\documentclass[
amsmath,amssymb,
]{revtex4-2}

\usepackage[T1]{fontenc}
\usepackage{graphicx}
\usepackage{amssymb}
\usepackage{amsthm}
\usepackage{amsmath}
\usepackage{mathtools}
\usepackage{algorithmic}
\usepackage{algorithm2e}

\usepackage{physics} 
\usepackage{comment}

\newcommand{\hilbert}[1]{H\left[{#1}\right]}
\newcommand{\dhilbert}[1]{H_d\left({#1}\right)}

\newcommand{\sgn}{{\rm sgn}}
\newtheorem*{theorem}{Theorem}

\begin{document}

\preprint{APS/123-QED}

\title{An extended Hilbert transform method for reconstructing the phase from an oscillatory signal}

\author{Akari Matsuki$^1$}
\author{Hiroshi Kori$^{1,2}$}
\author{Ryota Kobayashi$^{2,3,4}$}
\affiliation{
    $^1$Graduate School of Information Science and Technology, The University of Tokyo, Tokyo, 113-8656, Japan \\
    $^2$Graduate School of Frontier Sciences, The University of Tokyo, Chiba, 277-8561, Japan \\
    $^3$Mathematics and Informatics Center, The University of Tokyo, Tokyo, 113-8656, Japan \\
    $^4$JST PRESTO, Saitama, 332-0012, Japan
}
    

\keywords{Oscillation, Synchronization, Hilbert transform,  Bedrosian’s theorem}

\begin{abstract}
Rhythmic activity is ubiquitous in biological systems from the cellular to organism level. Reconstructing the instantaneous phase is the first step in analyzing the essential mechanism leading to a synchronization state from the observed signals. 
A popular method of phase reconstruction is based on the Hilbert transform, which can only reconstruct the interpretable phase from a limited class of signals, e.g., narrow band signals. 
To address this issue, we propose an extended Hilbert transform method that accurately reconstructs the phase from various oscillatory signals. 
The proposed method is developed by analyzing the reconstruction error of the Hilbert transform method with the aid of Bedrosian's theorem.
We validate the proposed method using synthetic data and show its systematically improved performance compared with the conventional Hilbert transform method with respect to accurately reconstructing the phase. Finally, we demonstrate that the proposed method is potentially useful for detecting the phase shift in an observed signal.
The proposed method is expected to facilitate the study of synchronization phenomena from experimental data.
\end{abstract}  

\maketitle

%
%
\thispagestyle{empty}


\section{Introduction}

Rhythmic activity is ubiquitous in biological systems, including cortical networks in the brain\cite{buzsaki2004neuronal,wang2010}, human heart and respiratory system\cite{schafer1998,lotrivc2000,kralemann2013vivo}, circadian rhythm\cite{fukuda2013controlling,yamaguchi2013}, gene expression\cite{yoshioka2020coupling}, and animal gait\cite{collins1993,borgius2014spinal,kobayashi2016}.
The phase description approach\cite{winfree1980,kuramoto1984chemical} describes the state of a multi-dimensional nonlinear oscillator using a variable called the phase and derives a reduced phase equation from a nonlinear dynamical system. 
This approach has promoted the understanding of how a population of nonlinear oscillatory elements can synchronize or form a cluster state. 
Theoretical studies based on the phase equation have been used to investigate the potential mechanisms underlying synchronization phenomena, including mutual coupling among the elements and the common inputs to the elements~\cite{ashwin2016,nakao2016}.

Fundamental questions in complex systems include how a system in the real-world achieves synchronization and what is the essential mechanism that leads to a synchronization state~\cite{pikovsky2003synchronization}. 
While the theoretical studies provide potential explanations for the synchronization phenomena, they cannot directly answer these questions. 
It is essential to reconstruct the instantaneous phase from observed data (e.g., signals or time series) and to infer the phase equation from the reconstructed phase. 
Many studies have focused on the latter step, that is, they have developed the inference methods for the phase response curve\cite{galan2005efficient,ota2009weighted,nakae2010bayesian,cestnik2018inferring,namura2022estimating} and the coupling function\cite{rosenblum2001detecting,tokuda2007inferring,kralemann2008phase,ren2010noise,levnajic2011network,stankovski2012inference,ostergaard2017oscillating,onojima2018dynamical,suzuki2018bayesian} from the phase (various reviews discuss this topic \cite{stankovski2017coupling,tokuda2019practical}). 
Conversely, a few studies\cite{gengel2019phase} have focused on the former step, i.e., the reconstruction of the instantaneous phase from an observed signal. 
An accurate phase reconstruction is necessary to study the synchronization phenomena in data because these inference methods assume the perfect phase reconstruction.

There are two primary approaches to reconstructing the instantaneous phase from an oscillatory signal. 
One simple approach to reconstructing the phase is to use linear interpolation between the subsequent marker events. 
For example, the phase is defined as 0 or $2\pi$ at the time of the action potential (spike) for neuronal oscillators\cite{galan2005efficient,ota2009weighted,nakae2010bayesian} or a heartbeat\cite{schafer1998}.  
This method can accurately reconstruct the phase when the noise level is not very high. However, this method is not applicable to signals without identifiable marker events, such as neuronal spikes. 
An alternative phase reconstruction approach is to apply the Hilbert transform to the observed signal\cite{gabor1946theory, pikovsky2003synchronization, king2009hilbert-vol1}. 
An advantage of the Hilbert transform method is that it is applicable even when there is no well-defined marker.  
Consequently, the Hilbert transform method has been applied to a variety of systems, e.g., the respiratory system in human\cite{schafer1998,kralemann2013vivo}, the gene expression in a cell\cite{yoshioka2020coupling}, and the human brain activity \cite{chavez2006towards,fujisawa20114,onojima2018dynamical,schreglmann2021non}. 
The limitation of the Hilbert transform method is that it can reconstruct the physically interpretable phase from a limited class of signals, i.e., the narrow band signals\cite{cohen1999ambiguity,chavez2006towards}. 
Therefore, it is necessary to carefully develop a pre-processing procedure via trial and error, which hinders the application of this method to oscillatory signals. 
Theoretical studies in signal processing have clarified the mathematical conditions of the signals on which the Hilbert transform method can reconstruct a meaningful phase\cite{delprat1992,cohen1999ambiguity,chavez2006towards}. 
However, only a few attempts have been made to develop a method to reconstructing the phase from more general signals. 

In this study, we propose an extension of the Hilbert transform method that can reconstruct the interpretable phase from a wider variety of signals. 
Here, we consider a new class of signals, called "weakly phase-modulated signals," which are an extension of the sinusoidal signals from which the conventional Hilbert transform method can reconstruct the phase. 
We first demonstrate that this conventional method cannot accurately extract the phase from these signals (Fig.~\ref{fig:HTdefect}). 
Then, we derive a new algorithm to reconstruct the phase from the phase-modulated signals and empirically show that the proposed method improves the reconstruction performance.  

This paper is organized as follows. 
We first review the conventional Hilbert transform method for reconstructing the instantaneous phase from data. In addition, we illustrate the limitation of the conventional method using an example. 
Second, we present the proposed method for reconstructing the instantaneous phase and examine the computational complexity of the algorithm. 
Third, we evaluate the performance of the phase reconstruction and compare its performance with that of the conventional method. Finally, we conclude this study and discuss future directions.

\section{Results}
\subsection{Estimating the instantaneous phase from an oscillatory signal}
\label{subsec:setting} 

A standard method for reconstructing the instantaneous phase from an oscillatory signal is based on the Hilbert Transform (HT)\cite{gabor1946theory, pikovsky2003synchronization, king2009hilbert-vol1}. 
This method calculates the phase from the analytic signal, defined as 
\begin{eqnarray}
    \zeta (t) = y(t) + i\hilbert{y(t)}, \label{eq:analytic}
\end{eqnarray}
where $y(t)$ and $\hilbert{y(t)}$ are the observed signal and its HT 
\begin{eqnarray}
    \hilbert{y(t)} = \pi^{-1} {\rm P.V.}\int_{-\infty}^{\infty} \frac{y(\tau)}{t-\tau} d\tau, \label{eq:HT-cont-def-int}
\end{eqnarray}
where ${\rm P.V.}$ refers to the Cauchy principal value. 
The HT method reconstructs the instantaneous phase by the argument of the analytic signal 
\begin{eqnarray}
    \phi^{\rm H}(t) = \arg \left[\zeta(t) \right]. \label{eq:phase_HT}
\end{eqnarray}
It is well-known that the HT method can reconstruct the interpretable phase from a particular class of signals. 
Let us consider the sinusoidal signal 
\begin{eqnarray}
    x(t)= A_0 \cos\left( \hat{\omega} t + \phi_0 \right),     
\end{eqnarray}
where $\hat{\omega}$ is the effective frequency, and $\phi_0$ is the initial phase. The HT method can perfectly reconstruct the interpretable phase from the signal: $\phi^{\rm H}(t)= \hat{\omega} t + \phi_0$. 
Furthermore, it is possible to extend this result to signals with slow amplitude modulation 
\begin{eqnarray}
    x(t)= A_L(t) \cos\left( \hat{\omega} t + \phi_0 \right), \label{eq:x-slow-amp}
\end{eqnarray}
where the amplitude $A_L(t)$ is the low-pass-filtered signal whose Fourier coefficients of the frequency higher than the effective frequency ($f> \hat{\omega}$) vanish. 
It can be shown~\cite{bedrosian1963product} that the HT method can perfectly reconstruct the phase: $\phi^{\rm H}(t)= \hat{\omega} t + \phi_0$. 
However, the HT method can only reconstruct the interpretable phase from a particular class of signals, i.e., the narrow band signals\cite{cohen1999ambiguity,chavez2006towards}.

In this study, we extend the HT method for 
another type of signal, which we call "weakly phase-modulated signals" 
\begin{equation}
    x(t)= A_0 \cos \left( \hat{\omega} t + \phi_0+ u(t) \right),   \label{eq:phase_mod}
\end{equation}
where $u(t)$ is the phase-modulation from the sinusoidal signal. We set $\phi_0=0$ without loss of generality by shifting the time. 

\begin{figure}[t]
\begin{center}
    \includegraphics[scale=.45]{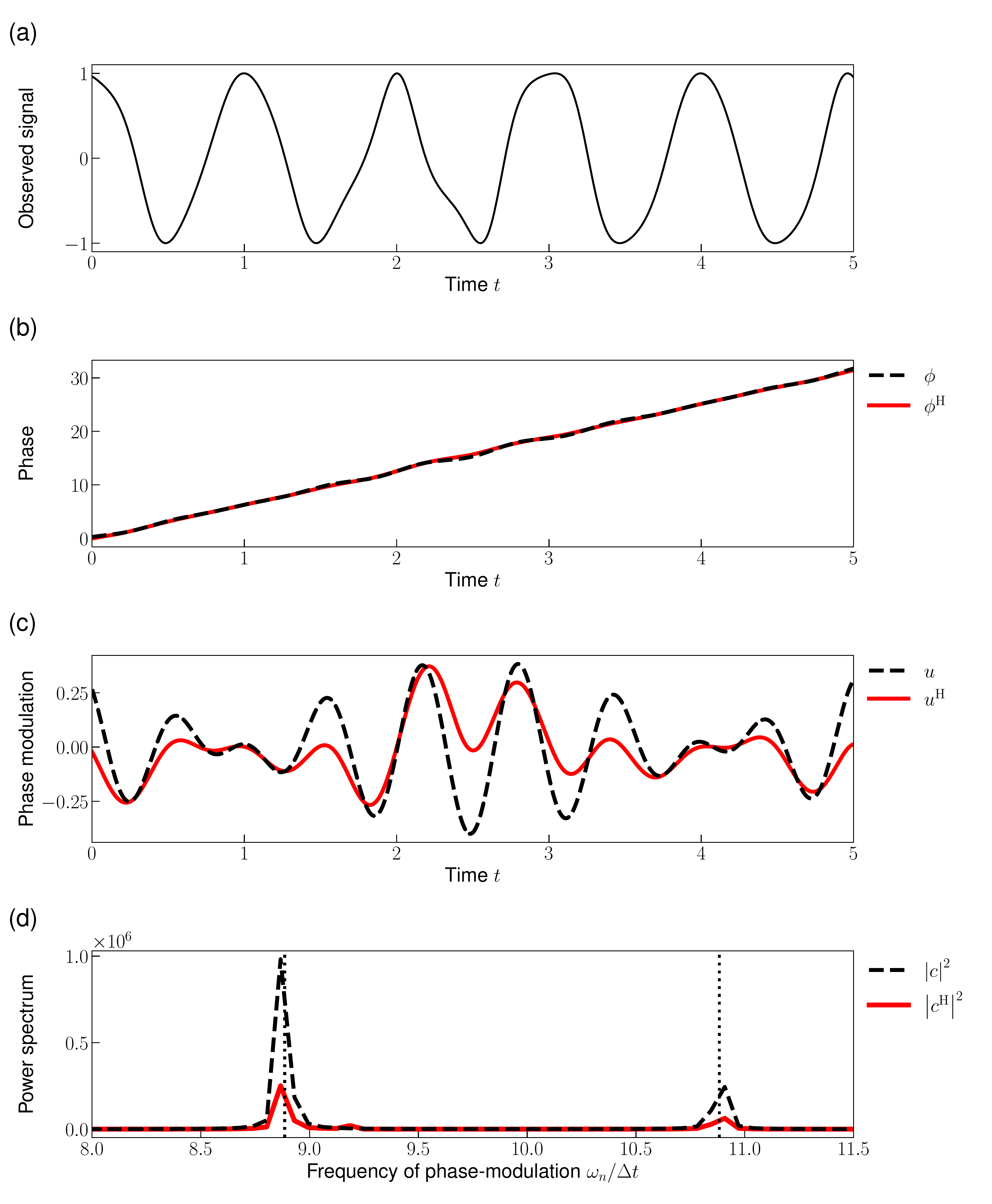} 
    \caption{Reconstruction of the instantaneous phase by using the Hilbert transform (HT) method. \\ 
    (a): Observed signal $x(t)$ (Eq. \ref{eq:phase_mod}) with $\phi_0= 0$ and $u(t) = 0.2 \left(\sin\sqrt{2} \hat{\omega} t + \cos\sqrt{3} \hat{\omega} t \right)$. 
    (b): Instantaneous phase $\phi(t)$.  
    (c): Phase-modulation $u(t)= \phi(t)- \hat{\omega} t- \phi_0$. 
    (d): Spectra density of the phase-modulation. 
    The dashed line in (b) and (c) represents the true phase and phase-modulation, respectively. 
    The red line in (b) and (c) represents the phase and phase-modulation reconstructed by the HT method, respectively.     
    Dotted vertical lines in (d) represent the dominant frequencies of the true phase-modulation: $\sqrt{2} \hat{\omega}$ and $\sqrt{3} \hat{\omega}$, where $\hat{\omega}= 2\pi$ is the effective frequency.
    }      \label{fig:HTdefect}
\end{center}
\end{figure}
\clearpage

We applied the HT method to a phase-modulated signal (Fig.~\ref{fig:HTdefect}(a)).
Figure~\ref{fig:HTdefect}(b) demonstrates that the HT method can accurately track the linear trend $\hat{\omega} t$ and estimate the effective frequency $\hat{\omega}$ even from a  phase-modulated signal. 
Note that this method (Fig.~\ref{fig:HTdefect}(c), red) cannot accurately reconstruct the phase-modulation $\phi(t)- \hat{\omega} t$. 
Then, we analyzed the power spectrum of the phase-modulation $\phi(t)- \hat{\omega} t$ to investigate the effect of the HT method. 
Figure~\ref{fig:HTdefect}(d) compares the power spectrum of the phase-modulation reconstructed using the HT method with that of the true phase-modulation. 
We plotted the frequency range of $\hat{\omega} < f < 2 \hat{\omega}$ because the phase-modulation is given by the sum of two sinusoidal functions ($\sin  \sqrt{2} \hat{\omega} t$ and $\cos \sqrt{3} \hat{\omega} t$) in this example. 
The result indicates that the HT method behaves like a low-pass filter, that is, it suppresses the spectral density of the peak frequencies ($f= \sqrt{2} \hat{\omega}, \sqrt{3} \hat{\omega}$). 
Motivated by this observation, we investigate how the HT method changes the power spectrum in the following subsection. 
Then we extend the HT method to reconstruct the instantaneous phase from an oscillatory signal; the proposed method preserves the power spectrum of the phase-modulation $u(t)$. 

Note that it is critical to reconstruct the phase-modulation $u(t)$ accurately to study the synchronization mechanism~\cite{pikovsky2003synchronization}, even though the modulation is small (Fig.~\ref{fig:HTdefect}(b), (c)). 
Many methods for inferring the phase coupling function rely on the assumption that the phase has been perfectly reconstructed; consequently, 
the bias in the phase reconstruction may induce a serious effects on the inference results.

\subsection{Proposed method}
As we observed in Fig.~\ref{fig:HTdefect}, the conventional HT method cannot reconstruct an interpretable phase from phase-modulated signals. In this subsection, we extend the HT method to include the phase-modulated signals (Eq. \ref{eq:phase_mod}).   

Let us assume that the signal is sampled at $N$ time steps with a constant interval $\Delta t$. 
We consider a phase modulated signal (Eq. \ref{eq:phase_mod}) sampled at time $t= k\Delta t$ 
\begin{equation}
    x[k]:= x(k\Delta t)= A_0 \cos \left( \phi[k] \right),   \label{des_phase_mod}
\end{equation}
where $A_0$ is the amplitude and $\phi[k] := \phi(k\Delta t)$ is the instantaneous phase at time $t= k\Delta t$.

We can analyze the effect of the phase-modulation on the phase reconstructed via the HT method with the aid of Bedrosian's theorem. 
The true phase-modulation $u[k]:= u(k\Delta t)$ and its reconstruction via the HT method $u^{\rm H}[k]:= \phi^{\rm H}(k\Delta t) - \hat{\omega} k\Delta t $ can be represented as Fourier series:  
\begin{equation}
    u[k]   = \sum_{n=0}^{N-1} c_n   e^{i k \omega_n},   \quad   
    u^{\rm H}[k] = \sum_{n=0}^{N-1} c^{\rm H}_n e^{i k \omega_n}, 
\end{equation}
where $\omega_n= 2n \pi/N$, 
and $c_n$ and $c^{\rm H}_n$ are given by the discrete Fourier transform of $u[k]$ and $u^{\rm H}[k]$, respectively. 
Assuming that the phase-modulation is small: $\epsilon:= \max_k  |u[k]| \ll 1$, we can derive a formula that clarifies the relation between the Fourier coefficients ($c_n$ and $c^{\rm H}_n$) (see Methods for the derivation), 
\begin{eqnarray}   \label{eq:Ch}
  c^{\rm H}_{n} \approx
        \begin{dcases}
            c_n - \frac{1}{2} \bar{c}_{2m-n} - \frac{1}{2} c_{n+2m} 
                & {\rm for} \quad  0 \le n \le m-1 ,  \\
            \frac{3}{4} c_m - \frac{1}{4} \bar{c}_{m} - \frac{1}{2} c_{3m}
                & {\rm for} \quad  n= m ,    \\
            \frac{1}{2} c_n - \frac{1}{2} c_{n+ 2m}  
                & {\rm for} \quad  m+1 \le n \le  N/2- 2m,     \\
            \frac{1}{2} c_n  
                & {\rm for} \quad  N/2- 2m + 1 \le n \le  N/2,    \\
        \end{dcases}
\end{eqnarray}
where $m:= \hat{\omega} N \Delta t/ 2\pi$ is a frequency index corresponding to the effective frequency
$\hat{\omega}$, and $\bar{z}$ denotes the complex conjugate of a complex number $z$. 
In addition, it is assumed that the observation is sufficiently long to satisfy $N > 4m$, and that the number of data points $N$ is even. If this number is odd, the term $N/2$ should be replaced with $(N-1)/2$. 

This result (Eq. \ref{eq:Ch}) illustrates the effect of the HT method on the phase-modulation in the frequency domain. 
Eq. (\ref{eq:Ch}) shows that the phase reconstructed by the conventional HT method is inconsistent with the true phase for phase-modulated signals.
This is because the Fourier coefficients reconstructed via the HT method $c^{\rm H}_n$ are not equal to those of the true phase-modulation $c_n$. 
In addition, the result (Eq. \ref{eq:Ch}) implies that the HT method acts as a low-pass-like filter to the phase-modulation $u(t)$. 
Let us consider the phase-modulated signal with a single frequency component $j$: 
\begin{eqnarray}
    c_n = \begin{cases}
        \alpha & {\rm for} \quad n= j,\\
        \bar{\alpha} & {\rm for} \quad n= N-j, \\
        0 & {\rm otherwise},
    \end{cases}     \nonumber   
\end{eqnarray}
where $\alpha$ is a non-zero complex value. 
If the phase-modulation frequency is lower than the effective frequency: $j < m$, the HT method perfectly reconstructs the true phase, i.e., $c^H_n= c_n$ for all the $n$. 
Conversely, when the phase-modulation frequency is higher than the effective frequency: $j> m$, the amplitude of the reconstructed phase-modulation is half of that of the true phase-modulation, i.e., $c^H_n= c_n/2$ for all the $n$. 
Indeed, Fig.~\ref{fig:HTdefect}(d) shows that the Fourier coefficient of the reconstructed phase-modulation $c^{\rm H}_n$ is smaller than the true modulation $c_n$ near the dominant Fourier modes ($\sqrt{2} \hat{\omega}$ and $\sqrt{3} \hat{\omega}$).

We can extend the HT method to accommodate phase-modulated signals. In the following, we describe the proposed method, which consists of five steps (Algorithm~\ref{alg:proposed-method}).  
First, we calculate the initial guess of the phase $\phi^{\rm H}[k]$ by using the conventional HT method (Eq. \ref{eq:phase_HT}). 
The Gibbs phenomenon dramatically impairs the phase reconstruction of the HT method when there is a large discrepancy between the values of the first and last point\cite{schreglmann2021non}. 
To mitigate this phenomenon, we extract the peaks from the signal and restricted the analysis to be from the first peak to the last one before applying the HT method.
Second, we estimate the effective frequency $\hat{\omega}$ from the initial guess: $\hat{\omega}= (\phi^{\rm H}[N-1]-\phi^{\rm H}[0])/T$, where $T= { (N-1) \Delta t }$ is the observation duration. 
Third, we calculate the discrete Fourier transform of the initial guess $\{ c^{\rm H}_n\}$ ($n= 1, 2, \dots, N$). 
Fourth, we correct the Fourier coefficient by inverting Eq. (\ref{eq:Ch}),  
\begin{eqnarray}   \label{eq:Inv_Ch}
  c^{\rm P}_n =
        \begin{dcases}
            2 c^{\rm H}_n  
                & {\rm for} \quad  N/2- 2m + 1  \le n \le  N/2 ,    \\
            2 c^{\rm H}_n - \frac{1}{2} \bar{c}^{\rm P}_{n+ 2m}  
                & {\rm for} \quad   m+1 \le n \le  N/2- 2m  ,     \\
            \Re{2 c^{\rm H}_n + c^{\rm P}_{3m} } + i \Im{ c^{\rm H}_n + \frac{1}{2} c^{\rm P}_{3m} } 
                & {\rm for} \quad  n= m ,    \\
            c^{\rm H}_n + \frac{1}{2} \bar{c}^{\rm P}_{2m-n} + \frac{1}{2} c^{\rm P}_{n+2m}
                & {\rm for}\quad   0 \le n \le m-1 ,  \\
        \end{dcases}
\end{eqnarray}
where $c^{\rm P}_n$ is the corrected Fourier coefficient. 
The remaining coefficients $c^{\rm P}_n$ 
($N/2 < n \leq N-1$) are calculated by using the formula $c^{\rm P}_{N-n}= \bar{c}^{\rm P}_n$ that reflects the fact that 
the phase-modulation
$u[k]$ is the real signal. 
Finally, in the fifth step, we reconstruct the phase-modulation by calculating the inverse Fourier transform of $\{ c^{\rm P}_n \}$ and smoothing the phase signal. 
We identify outliers in the reconstructed phase using Median Absolute Deviation criteria~\cite{leys2013detecting} and replace the outliers with a linear interpolation of the nearest neighbors.

\RestyleAlgo{ruled}
\begin{algorithm}[t]
    \caption{Proposed method for reconstructing the instantaneous phase.}
    \label{alg:proposed-method}
    \begin{algorithmic}[1]    
    \STATE  Calculate the initial guess of the phase $\phi^{\rm H}[k]$ by using the Hilbert transform method (Eq. \ref{eq:phase_HT}). 
    \STATE  Estimate the effective frequency $\hat{\omega}= (\phi^{\rm H}[N-1]-\phi^{\rm H}[0])/T$, where $T$ is the observation duration.
    \STATE  Calculate the discrete Fourier transform of the initial guess  
    $\{ c^{\rm H}_n\}$ ($n= 1, 2, \dots, N$).
    \STATE  Correct the Fourier coefficient $\{ c^{\rm P}_n\}$ ($n=1,2,\dots, N$) by using Eq. (\ref{eq:Inv_Ch}). 
    \STATE  Reconstruct the phase-modulation $\{ u[k] \}$ by calculating the inverse Fourier transform of $\{ c^{\rm P}_n\}$ and smoothing it.     
\end{algorithmic}
\end{algorithm}

Finally, we compare the computational complexities of the conventional HT method and the proposed method for reconstructing the phase of a signal. 
Computational complexity, that is, the dependency of the computational time on the data length, is critical when analyzing a long signal. 
Let us denote the number of data points of the signal as $N$. 
The computational complexity of the HT method is $O(N\log N)$, because we calculate the discrete Hilbert Transform (HT) by using the discrete Fourier transform (See Method). 
Next, we evaluate the computational complexity of the proposed method. 
First, the proposed method computes the HT: $O(N\log N)$ (Step 1 in Algorithm~\ref{alg:proposed-method}). 
Next, the effective frequency is calculated: $O(1)$ (Step 2). 
Then, the discrete Fourier transform is computed: $O(N\log N)$ (Step 3) and the coefficients of the Fourier transform are corrected: $O(N)$ (Step 4). 
Finally, the method reconstructs the phase-modulation by calculating the inverse Fourier transform: $O(N\log N)$ and smoothing it: $O(N\log N)$ (Step 5). 
Therefore, the computational complexity of the proposed algorithm is $O(N\log N)$, which is comparable to the conventional method. 
\clearpage

\subsection{Reconstruction performance of the proposed method}
Here, we examine whether the proposed method can accurately reconstruct the instantaneous phase from an observed signal. 
First, we consider an oscillatory signal with a constant amplitude 
\begin{equation}
    x(t)= \cos \left( \hat{\omega} t + u(t) \right),   \label{eq:sig_1}
\end{equation}
where $\hat{\omega}$ is the effective frequency and $u(t)$ is the phase-modulation. The sampling time interval and the duration of the simulation are $\Delta t =0.01$ and $T=200$, respectively, unless otherwise stated. 

We evaluated the performance of the phase reconstruction by analyzing the synthetic data based on two types of phase-modulated signals. 
The first signal is a quasi-periodic phase-modulation,  
\begin{eqnarray}    
    u(t) = b \left( \sin \sqrt{2} \hat{\omega} t + \cos \sqrt{3} \hat{\omega} t \right),
    \label{eq:u-periodic}
\end{eqnarray}
where $b$ is the amplitude of the phase-modulation. 
The second signal is the Ornstein-Uhalenbeck (OU) type phase-modulation 
\begin{eqnarray}
    \frac{du}{dt} = -k u(t) + \sigma \eta(t), \label{eq:u-OU}
\end{eqnarray}
where $\eta(t)$ is the Gaussian white noise with zero mean and unit variance. 

\begin{figure}
\begin{center}
    \includegraphics[scale=.45]{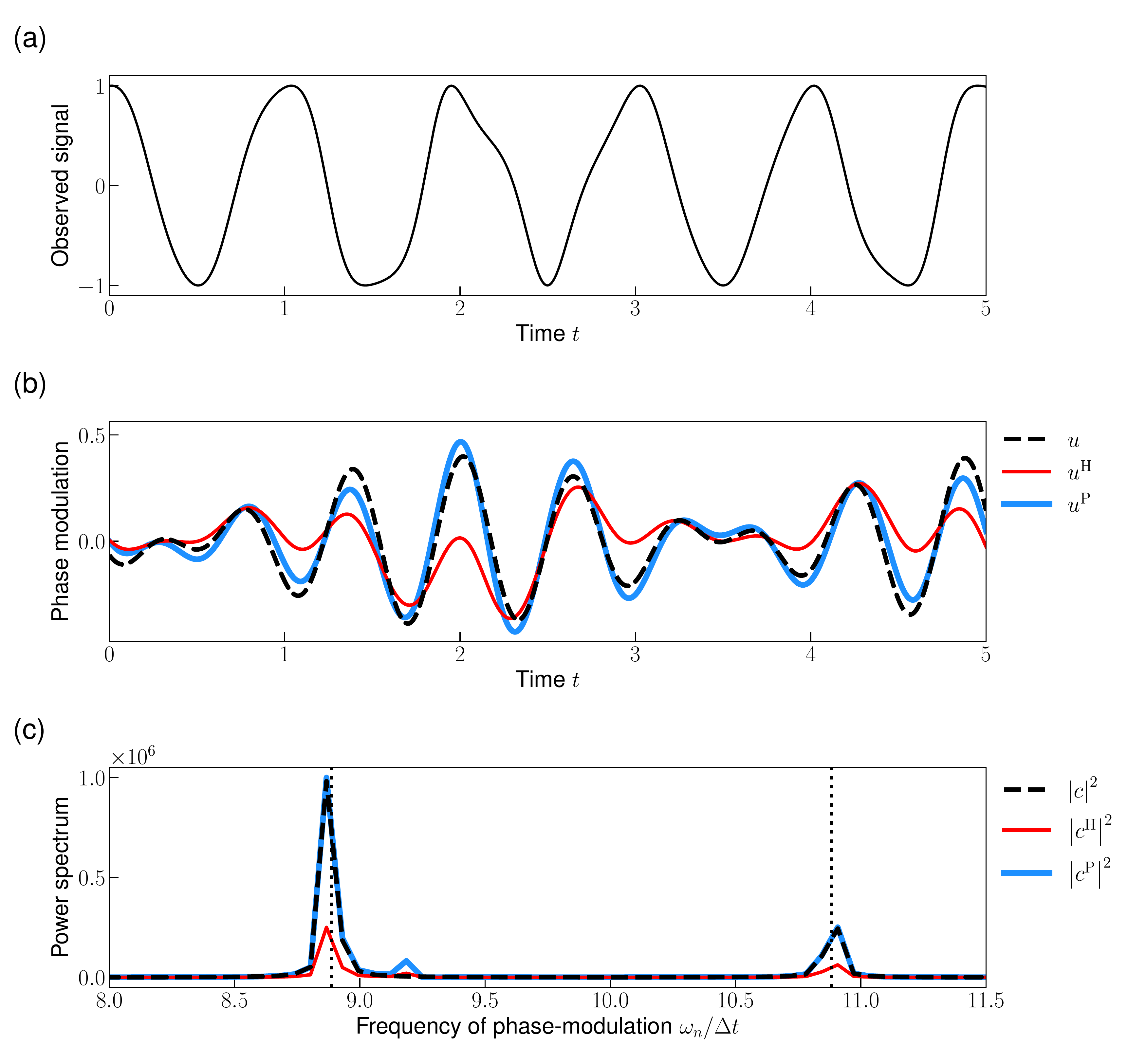}    
    \caption{Reconstructing the instantaneous phase by the proposed method: a signal with the quasi-periodic phase-modulation. \\
    (a): Observed signal $x(t)$ given by Eqs. (\ref{eq:sig_1}) and (\ref{eq:u-periodic}). 
    (b): phase-modulation $u(t)$. 
    (c): Power spectrum of the phase-modulation. 
    Dashed lines represent the true phase-modulation $u(t)$ in (b) and its power spectrum in (c). Red and blue lines represent the reconstructions by the conventional HT method and the proposed method, respectively. 
    The dotted vertical lines in (c) represent the dominant frequencies of the phase-modulation: $\sqrt{2} \hat{\omega}$ and $\sqrt{3} \hat{\omega}$. 
    Parameters are $\hat{\omega}=2\pi$ and $b=0.2$. 
    }
    \label{fig:estimation-periodic}
\end{center}
\end{figure}

\begin{figure}
\begin{center}
    \includegraphics[scale=.45]{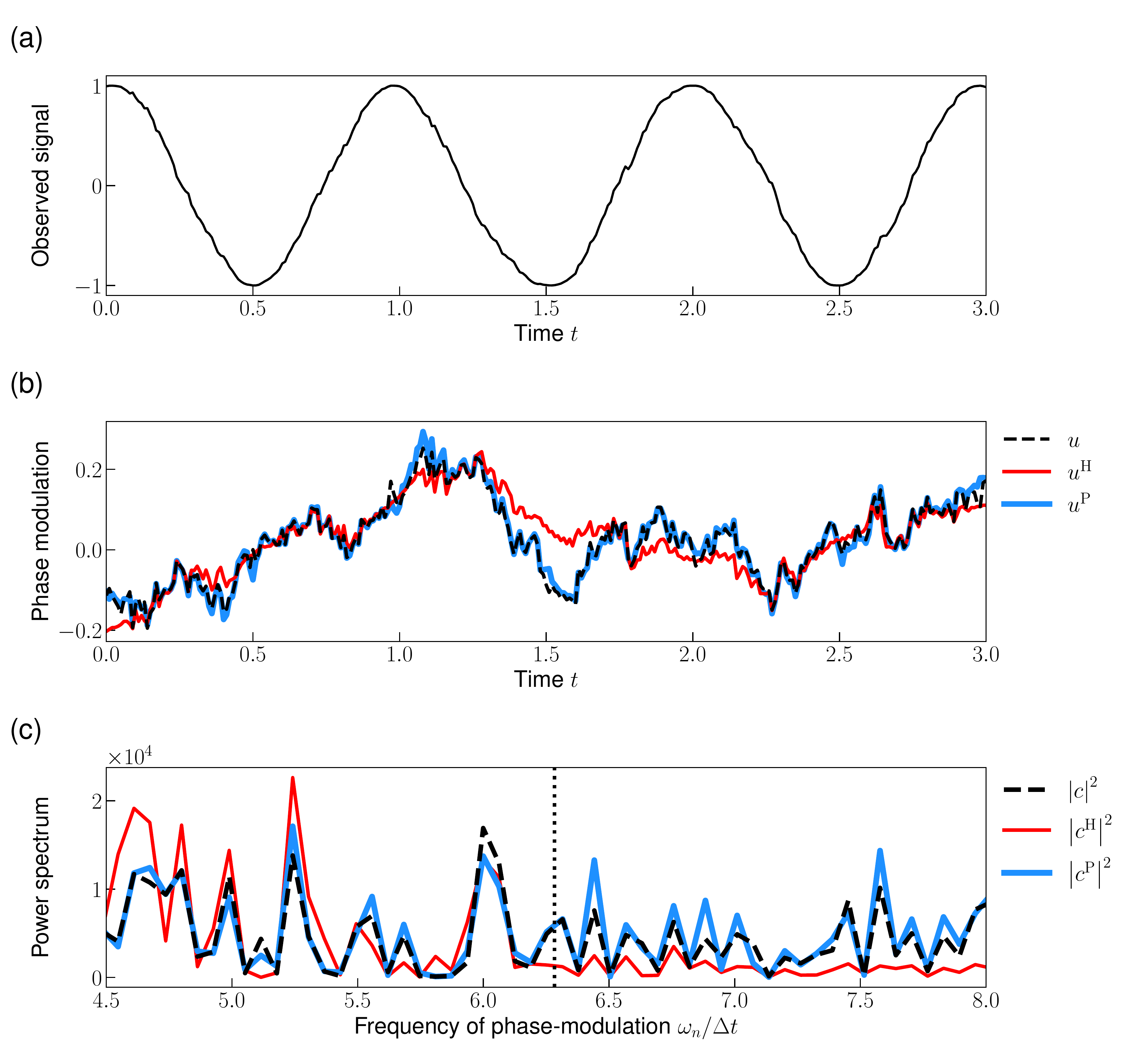}     
    \caption{Reconstructing the instantaneous phase by the proposed method: a signal with the OU-type phase-modulation.\\ 
    (a): Observed signal $x(t)$ given by Eqs. (\ref{eq:sig_1}) and (\ref{eq:u-OU}).          
    (b): phase-modulation $u(t)$. 
    (c): Power spectrum of the phase-modulation. 
    Dashed lines represent the true phase-modulation $u(t)$ in (b) and its power spectrum in (c). Red and blue lines represent the reconstructions by the conventional HT method and the proposed method, respectively.     
    The dotted vertical line in (c) represent the effective frequency $\hat{\omega}=2\pi$. Parameters are $k=2.0$, and $\sigma^2= 0.8$. 
    }    
    \label{fig:estimation-OU}
\end{center}
\end{figure}

We applied the proposed method to a signal with quasi-periodic phase-modulation (Fig.\ref{fig:estimation-periodic}(a)). 
Figure\ref{fig:estimation-periodic}(b) compares the phase reconstructed via the proposed method (blue) with that reconstructed via the conventional HT method (red). While the proposed method accurately reconstructs the phase-modulation, the conventional method cannot reconstruct it.  
In addition, we compared the power spectrum of the phase-modulation with that of the reconstructed phase-modulations (Fig.\ref{fig:estimation-periodic}(c)). 
We found that the proposed method can reconstruct a phase-modulation whose power spectrum is consistent with the true power spectrum. 
Next, we applied the proposed method to a signal with the OU-type phase-modulation (Fig.\ref{fig:estimation-OU}(a)). 
Similar to the case of the quasi-periodic modulation, the proposed method can reconstruct the phase-modulation (Fig.\ref{fig:estimation-OU}(b)) and its power spectrum (Fig.\ref{fig:estimation-OU}(c)) accurately. 
While the conventional HT method can track the slow trend of the phase fluctuation, it cannot accurately reconstruct the phase-modulation. 

\begin{figure}
\begin{center}
    \includegraphics[scale=.45]{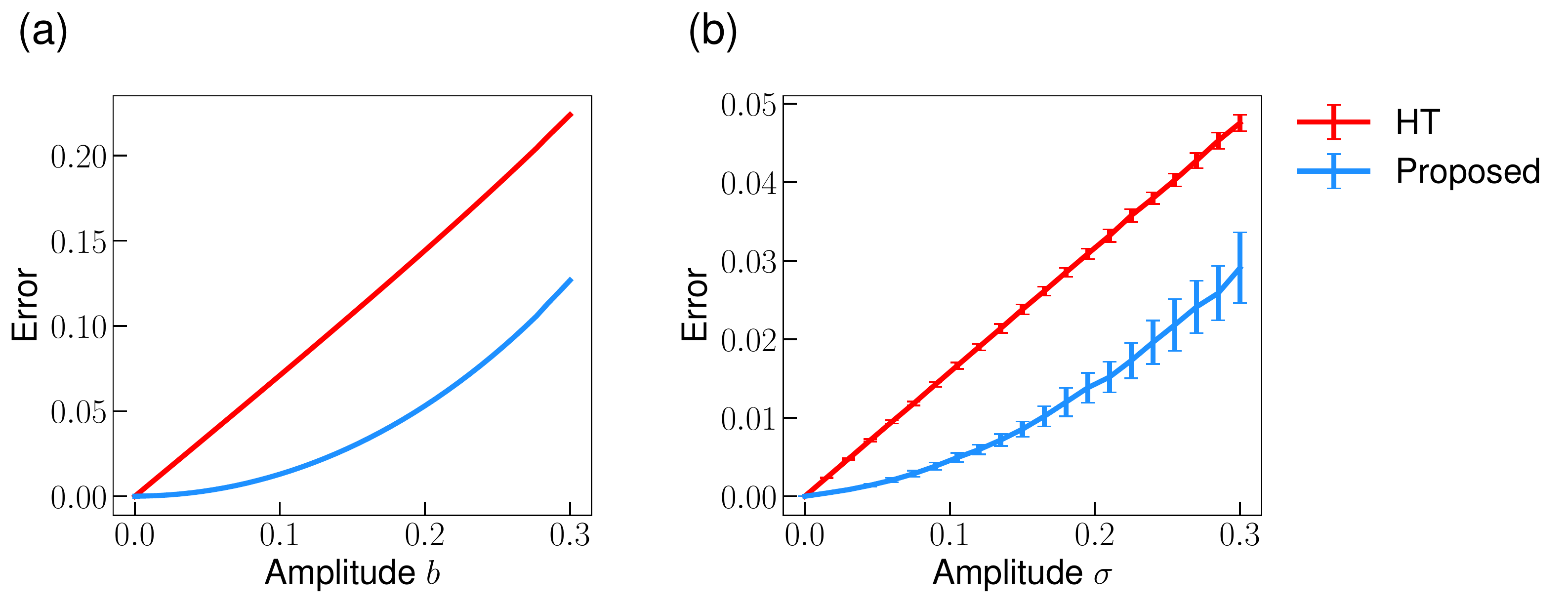}   
    \caption{Effect of the amplitude of phase-modulation on the phase reconstruction error. \\
    (a) Quasi-periodic phase-modulation (Eq. \ref{eq:u-periodic}). 
    (b) OU-type phase-modulation (Eq. \ref{eq:u-OU}). 
    We plotted the mean and standard deviation of the errors calculated from 100 trials in (b). Parameters are set as $\hat{\omega}= 2\pi$ and $k=2.0$ in (b). }
    \label{fig:error}
\end{center}
\end{figure}

Furthermore, we examined whether the proposed method can reconstruct the phase given a larger phase-modulation. 
We quantified the phase reconstruction performance based on the mean squared error of the phase-modulation. 
Figure \ref{fig:error}(a) shows that the proposed method consistently performs better than the conventional HT method for a signal with a quasi-periodic phase-modulation for a range of the phase-modulation amplitude $b$.
Nevertheless, the error of the proposed method increases with increasing phase-modulation amplitude. 
The performance deterioration may be due to the nonlinear effects, i.e., $O\left( \epsilon^2\right)$, which we neglected in the derivation of the method. 
Similarly, the proposed method consistently performs better than the HT method for a signal with an OU-type phase-modulation even when the phase-modulation is not small (Fig.\ref{fig:error} (b)). 
Even though we assumed a small phase-modulation to derive the proposed method, the results (Fig.\ref{fig:error}) suggests that the proposed method provides better performance than the HT method for signals with moderate phase-modulations. 
\clearpage

\begin{figure}
\begin{center}
    \includegraphics[scale=.45]{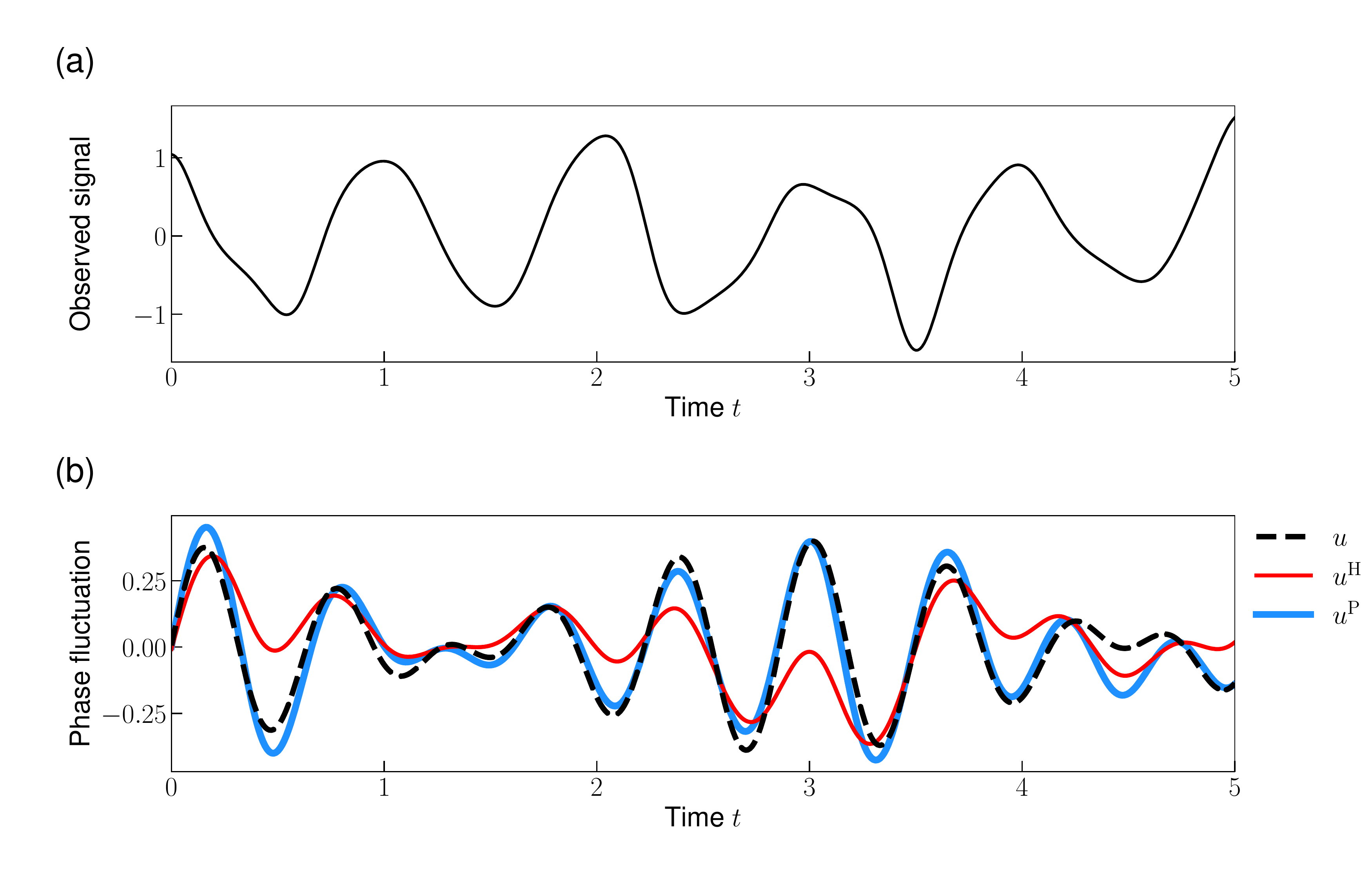}
    \caption{Reconstructing the instantaneous phase by the proposed method: a signal with amplitude and phase-modulation. \\ 
    (a) Observed signal $x(t)$ (Eq. \ref{eq:sig_2}). 
    (b) phase-modulation $u(t)$. 
    The dashed line represents the true phase-modulation $u(t)$, and the red and blue line represents its reconstruction by the HT method and the proposed method, respectively. 
    }            
    \label{fig:robustness}
\end{center}
\end{figure}

Finally, we examined whether the proposed method can reconstruct the phase from a signal with the amplitude and phase modulations. 
We consider a signal whose amplitude is modulated by a sum of sinusoidal functions
\begin{equation}
    x(t)= A(t) \cos \left( \hat{\omega} t + u(t) \right),   \label{eq:sig_2}
\end{equation}
where $\hat{\omega}$ is the effective frequency, 
\[ 
A(t) = 1+ 0.2\left(\cos 0.6 \hat{\omega} t + \sin 0.7 \hat{\omega} t \right), 
u(t)= 0.2 \left( \sin \sqrt{2} \hat{\omega} t+ \cos \sqrt{3} \hat{\omega} t \right).
\]
Figure \ref{fig:robustness} demonstrates that while the proposed method can accurately reconstruct the instantaneous phase from the amplitude and phase modulated signal, the conventional HT method cannot. 
This result can be understood as follows. 
Bedrosian's theorem~\cite{bedrosian1963product} states that the HT of the product of a high-pass and a low-pass signal with non-overlapping spectra is equal to that of the low-pass signal, and the HT of the high-pass signal. 
This theorem implies that the slow (or low-pass filtered) amplitude modulation (Eq.~\ref{eq:x-slow-amp}) will not impair the phase reconstruction by the conventional HT method. 
Thus, it is natural to expect that the proposed method works even when we observe the weakly phase-modulated signals. 

\begin{figure}
\begin{center}
    \includegraphics[scale=.45]{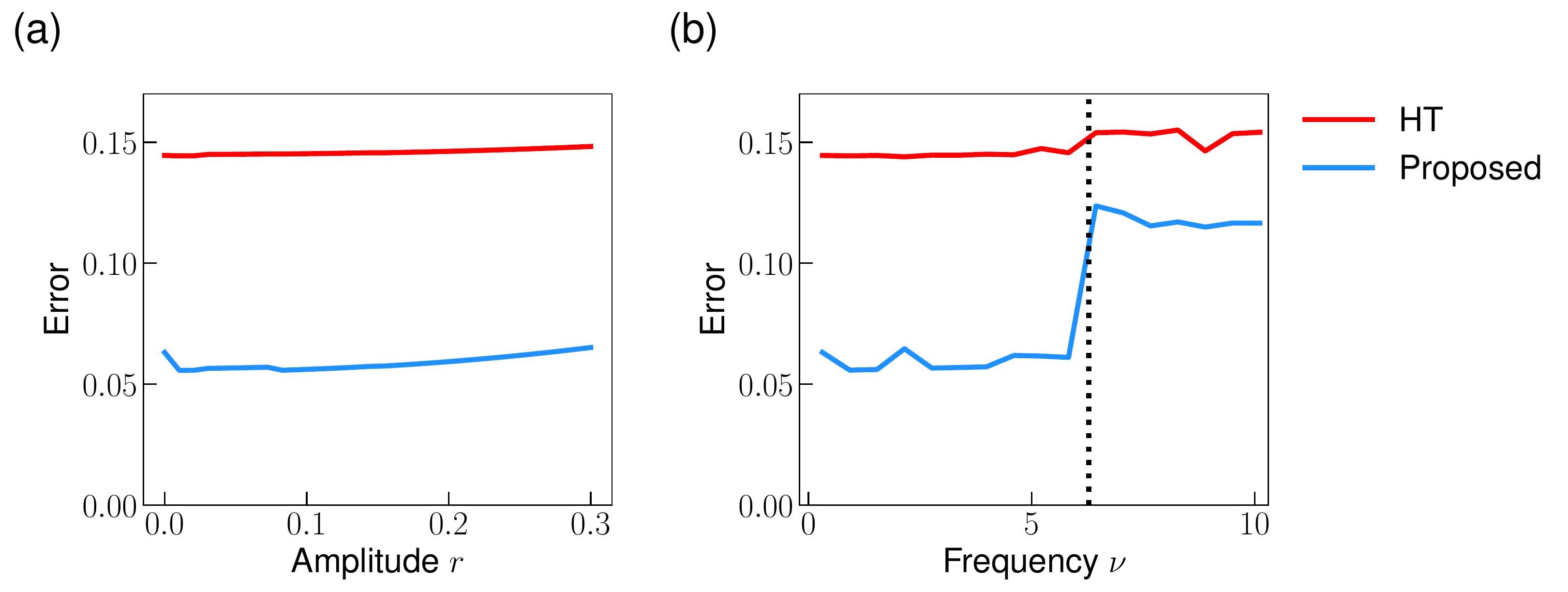}
    \caption{Effect of the amplitude modulation on the phase reconstruction error. \\     
    Dependence of the error on the amplitude $r$ (a) and the frequency $\nu$ (b) of the amplitude modulation.  
    Parameter were set as $\nu= 4.4$ in (a) and $r= 0.1$ in (b). The dotted vertical line in (b) represents the effective frequency $\hat{\omega}= 2 \pi$. }
    \label{fig:Error_AmpMod}
\end{center}
\end{figure}

Furthermore, we examined whether the proposed method is robust against the fast amplitude modulation. 
We consider an amplitude and phase modulated signal (\ref{eq:sig_2}) with 
\[  A(t) = 1+ r \cos\nu t, \quad  
    u(t)= 0.2 \left( \sin \sqrt{2} \hat{\omega} t + \cos \sqrt{3} \hat{\omega} t \right). \] 
Figure~\ref{fig:Error_AmpMod} shows the dependence of the reconstruction error on the amplitude $r$ and frequency $\nu$ of the amplitude modulation. 
The error does not depend on the amplitude, which indicates that the proposed method works even for signals with moderate amplitude modulation (Fig.~\ref{fig:Error_AmpMod}(a) ). 
While the error does not depend on the frequency $\nu$ in the range of $\nu < \hat{\omega}$ (Fig.~\ref{fig:Error_AmpMod}(b) ), it increases when the frequency becomes larger than the effective frequency $\hat{\omega}$. Nevertheless, the error of the proposed method is smaller than the HT method. 
Overall, these results suggest that the proposed method improves the phase reconstruction even for the signals with amplitude modulation.

\subsection{Detecting a phase shift from an observed signal} 
Biological oscillatory systems often exhibit "phase shifts,"  that is, a rapid change in the phase of a rhythm. 
For example, the phase of a circadian rhythm can change as a result of light exposure~\cite{vitaterna2001overview}. It would be useful to develop a method for detecting the phase shifts in oscillatory signals. 
We examined whether the proposed method is potentially useful for detecting the phase shifts in data. As a minimal model, we consider a single oscillator exhibiting a phase shift:
\begin{eqnarray}
    x(t) = \cos \left( \hat{\omega}_1 + u(t) \right),  \quad 
    \frac{du(t)}{dt} = \begin{cases}
        \sigma \eta(t) & {\rm for} \quad  t \notin [T_c, T_c+\Delta_T], \\
        \left( \hat{\omega}_2- \hat{\omega}_1 \right)+ \sigma \eta(t) & {\rm for} \quad  t \in [T_c, T_c+\Delta_T],
    \end{cases}     
    \label{eq:u-demo}
\end{eqnarray}
where the interval $[T_c, T_c+\Delta T]$ represents the change period, i.e., the period in which the phase of the oscillator shifts with a frequency $\hat{\omega}_2$, and $\eta(t)$ is the Gaussian white noise with zero mean and unit variance. 
The synthetic data were simulated with a total duration $T= 10$ and a change time $T_c= 5.0$.  

\begin{figure}
\begin{center}
    \includegraphics[scale=.45]{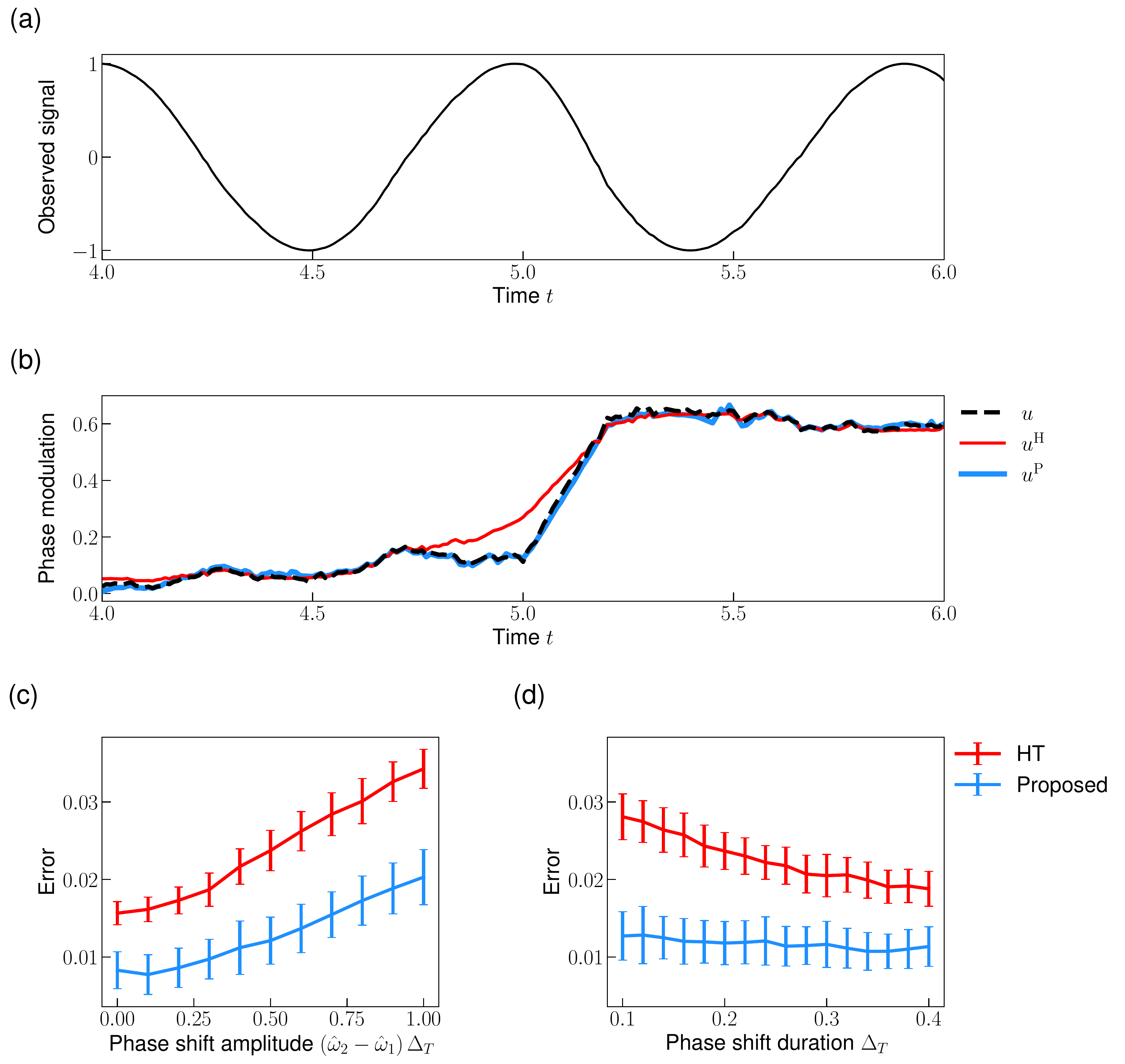}
    \caption{Detecting the phase shift from an oscillatory signal.\\ 
    (a): Observed signal $x(t)$ (Eq.~\ref{eq:u-demo}). 
    (b): phase-modulation $u(t)$. 
    (c, d): Dependence of the phase reconstruction error on the phase shift amplitude in (c) and the phase shift duration in (d).
    We plotted the mean and standard deviation of the errors calculated from 100 trials in (c) and (d).
    Parameters were set as the effective frequency $\hat{\omega}_1= 2\pi$, the noise variance $\sigma= 0.1$, 
    the shift amplitude $(\hat{\omega}_2- \hat{\omega}_1) \Delta_T= 0.5$ in (a), (b), and (d), and the 
    the shift duration $\Delta_T = 0.2$ in (a), (b), and (c). }      
    \label{fig:demo}
\end{center}
\end{figure}

Figure~\ref{fig:demo}(a) shows an observed signal before and after the change period, with the frequency increasing after $t= 5.0$. 
The proposed method can accurately track the change in the phase-modulation induced by the phase shift (Fig.~\ref{fig:demo}(b): Blue). 
Conversely, it is difficult for the conventional HT method to infer the change period because of the smoothing effect in the reconstructed phase (Fig.~\ref{fig:demo}(b): Red). 
Furthermore, we compared the phase reconstruction error of the proposed method with that of the conventional HT method. 
The proposed method achieved a smaller error than the HT method across a range of phase shift amplitude $(\hat{\omega}_2- \hat{\omega}_1) \Delta_T$ (Fig.\ref{fig:demo}(c)). 
Similar to the previous result concerning the reconstruction error (Fig.\ref{fig:error}), the errors of these methods increase with increasing phase shift amplitude. 
In addition, we examined the dependency of the error on the duration of the phase shift $\Delta_T$ when the phase shift amplitude is fixed. 
The error of the HT method increases as the shift duration decreases (Fig.\ref{fig:demo}(d): Red). 
Conversely, the error of the proposed method is small even for a signal with a small duration $\Delta_T$ (Fig.\ref{fig:demo}(d): Blue). 
This result suggests that the proposed method is more suitable for detecting rapid phase shifts than the HT method.

\section{Discussion}
We proposed an extension of the Hilbert Transform (HT) method for reconstructing the phase from an observed signal. 
We addressed a limitation of the conventional HT method, that is, the conventional method has been proven to work only for the narrow band signals.
We demonstrated that the conventional HT method cannot accurately reconstruct the interpretable phase from phase-modulated signals. Conversely, our method can extract the phase from these types of signals (Figs. \ref{fig:estimation-periodic}--\ref{fig:error}). 
Furthermore, we have demonstrated the performance of the proposed method by using the simulated data with the amplitude- and phase-modulated signals (Figs.~\ref{fig:robustness} and \ref{fig:Error_AmpMod}). 
Consequently, the extended HT method is a promising tool for investigating synchronization phenomena through analyzing oscillatory signals in biological systems. 

There are limitations to the proposed method, offering the opportunity for future improvements. 
First, we assumed that the phase-modulation $u(t)$ is so small that higher-order terms are negligible. 
In addition, we assumed that the amplitude of the signal is approximately constant. 
Despite of these assumptions, the numerical results (Figs.~\ref{fig:error}, \ref{fig:robustness} and \ref{fig:Error_AmpMod}) indicate that the proposed method is, to some extent, robust against violations of these assumptions. 
As shown in Fig. \ref{fig:Error_AmpMod}, an amplitude modulation of higher frequency can impair the performance of the proposed method. 
Future work is required to develop a phase reconstruction method that is robust to rapid amplitude modulations. 
Next, our method assumed a specific form of the observation signals, $s(t)= A(t) \cos (\hat{\omega} t+ u(t) )$. 
This assumption can fail in real-world systems. For example, the signals generated by the van der Pol oscillator\cite{namura2022estimating} and signals from ECG measurement\cite{kralemann2008phase} are distorted from the sinusoidal oscillations. To address this issue, a transformation method from the protophase to phase~\cite{kralemann2008phase, gengel2022phase} might improve the performance of the proposed method. 


Another future research direction is to apply the proposed method to a real-world dataset. 
We have shown that the proposed method can accurately reconstruct the high-frequency component compared with the conventional HT method. Therefore, the proposed method should be able to improve the estimation performances of the phase equations from the observed signals in biological systems. 
Moreover, the proposed method can identify the phase shifts in oscillatory signals more accurately than the conventional method (Fig.~\ref{fig:demo}). It would be an interesting future study to identify the phase shifts in the circadian rhythms using the proposed method. 

%
%


\section{Method}
\subsection{Discrete Hilbert transform}
We used the Hilbert Transform (HT) for discrete signals, i.e., we used the discrete Hilbert Transform (HT) to analyze the signals. 
In this subsection, we describe the definition and properties of the discrete HT. 
Let $X(\omega)$ be the Fourier transform of a continuous signal $x(t)$: 
\begin{eqnarray}
    x(t) = \int_{-\infty}^{\infty} d\omega X(\omega) e^{i\omega t}.
\end{eqnarray} 
The HT of $x(t)$ (Eq.~\ref{eq:HT-cont-def-int}) can be written by using its Fourier transform\cite{king2009hilbert-vol1} 
\begin{eqnarray}
    \hilbert{x(t)} = \int_{-\infty}^{\infty} d\omega \left( -i \cdot \sgn (\omega)\right)  X(\omega) e^{i\omega t}, \label{eq:ht-ft}
\end{eqnarray}
where $\sgn$ denotes the sign function defined as
\begin{eqnarray}
    \sgn (x) =
    \begin{cases}
        -1 & {\rm for} \quad x<0, \\
        0 & {\rm for} \quad x=0, \\
        1 & {\rm for} \quad x>0.
    \end{cases}
\end{eqnarray}
The discrete HT is defined to satisfy the property similar to Eq.(\ref{eq:ht-ft}). 
We consider the inverse discrete-time Fourier transform of 
a signal (sequence) $x[k]$: 
\begin{eqnarray}
    x[k] = \sum_{n=0}^{N-1} X_n e^{ik\omega_n},
\end{eqnarray}
where $X_n$ is the discrete Fourier transform of $x[k]$, $\omega_n= 2n \pi/N$ is the $n$-th frequency. 
The discrete HT of $x[k]$ is defined as 
\begin{eqnarray}
    \dhilbert{x[k]} 
    = \sum_{n=0}^{N-1} (-i S_n) X_n e^{ik\omega_n}, \label{eq:def-dHT1}
\end{eqnarray}
where $S_n$ is 
\begin{eqnarray}
    S_n = \begin{cases}
        0 & {\rm for } \quad n=0,\\
        1 & {\rm for } \quad 0 < n \le N/2, \\
        -1 & {\rm for } \quad N/2 < n \le N-1.
    \end{cases}     \label{eq:def-dHT2}
\end{eqnarray} 
Again, we assume that the number of data points $N$ is even. 
If this number is odd, the term $N/2$ should be replaced with $(N-1)/2$. 
From this definition (Eqs.\ref{eq:def-dHT1} and \ref{eq:def-dHT2}), we can derive the formula implying that the discrete HT can also reconstruct the phase of the sinusoidal wave,  
\begin{eqnarray}
    \dhilbert{ \cos(\omega k) }= \sin(\omega k),   \qquad 
    \dhilbert{ \sin(\omega k) }= -\cos(\omega k).
    \label{eq:HT-prop-sin}
\end{eqnarray}
where $0< \omega < \pi$ is the frequency parameter.

Next, we introduce Bedrosian's theorem, which states that 
the HT of the product of a high-pass signal and a low-pass signal with non-overlapping spectra is given by the product of the low-pass signal and the HT of the high-pass signal. Formally, it is written as follows: 
\begin{theorem}	[Bedrosian's theorem for the discrete Hilbert transform \cite{li2010discrete, king2009hilbert-vol1}] 
Let $x[k]$ and $y[k]$ ($k=0,1,\dots, N-1$) be sequences with their Fourier transform $X_n$ and $Y_n$  ($n=0,1,\dots, N-1$), respectively.
If there exists an integer $0 < m < N/2$ such that
\begin{eqnarray}
    \begin{array}{cl}
        X_n = 0 &\qquad {\rm for} \quad m \le n \le N-m, \\
        Y_n = 0 &\qquad {\rm for} \quad  0 \le n \le m-1,\ \ N-m+1 \le n \le N-1, 
    \end{array}
\end{eqnarray}
then the discrete Hilbert transform of the product of $x[k]$ and $y[k]$ is written as 
\begin{eqnarray}
    \dhilbert{x[k] y[k]} = x[k] \dhilbert{y[k]}.    \label{eq:Bedrosian_Th}
\end{eqnarray}
\end{theorem}

\subsection{Analysis of phase-modulation reconstructed via the conventional Hilbert Transform method}
Here, we analyze the phase-modulation reconstructed via the conventional HT method. The aim of this subsection is to derive a formula (Eq. \ref{eq:Ch}) that characterizes the relationship between the phase-modulation and its reconstruction.
Let us assume that we observe a weakly phase-modulated signal 
\begin{eqnarray}
    x[k] = \cos\phi[k],
    \label{eq:Dist_PMsignal}
\end{eqnarray}
where $\phi[k]= \hat{\omega} k\Delta t  + u[k]$ is the instantaneous phase at time $k\Delta t$,
$\hat{\omega}$ is the effective frequency, 
and $u[k]$ is a small phase-modulation: $\epsilon =\max_k \left|u[k]\right| \ll 1$. 
Note that the amplitude can be set as 1 ($A_0= 1$) without loss of generality.

The HT method reconstructs the phase via the argument of the analytic signal (Eq.~\ref{eq:phase_HT}). 
Substituting Eq.(\ref{eq:Dist_PMsignal}) into Eq.(\ref{eq:phase_HT}), we obtain 
\begin{eqnarray}
	\phi^{\rm H}[k]	
    &=& \arg \left[\ 
            \cos \phi[k] + i \dhilbert{  \cos \phi[k]  } 
            \ \right]
    =    \arg \left[\             
            \cos\left( \hat{\omega} k\Delta t+  u[k]  \right) 
               + i \dhilbert{  \cos\left( \hat{\omega} k\Delta t + u[k]  \right) } 
             \ \right] \nonumber  \\ 
    &\approx&   \arg \left[\ 
            \cos ( \hat{\omega} k\Delta t )
            - u[k] \sin ( \hat{\omega} k\Delta t )
               + i \dhilbert{  
               \cos ( \hat{\omega} k\Delta t )
                - u[k] \sin ( \hat{\omega} k \Delta t )    } 
             \ \right] \nonumber  \\ 
    &=& \arg \left[  \left\{ \cos(\hat{\omega} k\Delta t)+             i  \sin(\hat{\omega} k\Delta t)  \right\} -  
              \left\{  u[k] \sin(\hat{\omega} k\Delta t) +     i H_d( u[k] \sin(\hat{\omega} k\Delta t) )    \right\}  \right] \nonumber  \\    
	&=& \hat{\omega} k\Delta t + 
        \arg \left[ 1 - e^{-i \hat{\omega} k \Delta t}
            \left\{ u[k] \sin( \hat{\omega} k \Delta t) + i\dhilbert{u[k] \sin( \hat{\omega} k \Delta t) } \right\} 
        \right] \nonumber\\
	&\approx& \hat{\omega} k\Delta t -
        {\rm Im} \left[ e^{-i\hat{\omega}k\Delta t} 
            \left\{ u[k] \sin( \hat{\omega}k\Delta t )+ 
            i \dhilbert{u[k] \sin( \hat{\omega}k\Delta t) } \right\} 
        \right], 
	\label{eq:estimator_HT}
\end{eqnarray}
where the approximation symbol $\approx$ represents that the higher order terms $O(\epsilon^2)$ are neglected, and ${\rm Im}[z]$ denotes the imaginary part of a complex number $z$. 
Hence, the phase-modulation reconstructed via the conventional HT method $u^{\rm H}[k]:=\phi^{\rm H}[k] - \hat{\omega}k\Delta t$ can be written as follows: 
\begin{eqnarray}
    u^{\rm H}[k] \approx f\left( u[k] \right) 
    := -{\rm Im} \left[ e^{-i\hat{\omega}k\Delta t} \left\{ u[k] \sin( \hat{\omega}k\Delta t )+ i \dhilbert{u[k] \sin( \hat{\omega}k\Delta t) }\right\} \right]. \label{eq:def-f}
\end{eqnarray}
To analyze the discrete HT in Eq.(\ref{eq:def-f}), we consider the discrete Fourier series of the phase-modulation
\begin{eqnarray}
    u[k] = \sum_{n= 0}^{N-1}  c_n e^{i k \omega_n}  
    = \sum_{n=0}^{N/2} v_n[k],  \label{eq:u-vn}         
\end{eqnarray}
where $\omega_n= 2 \pi n/N$ is a frequency of the $n$-th Fourier component, 
\begin{eqnarray}    
    v_n[k] = \begin{cases}
        c_n e^{ik\omega_n} & {\rm for} \quad  n= 0 \quad {\rm or} \quad n= N/2,  \\
         c_n e^{ik\omega_n} + \bar{c}_n e^{-ik\omega_n} & {\rm otherwise}, 
    \end{cases}    \nonumber 
\end{eqnarray}
is the $n$-th frequency component. 
For the derivation of Eq. (\ref{eq:u-vn}), we used the formula $c_{N-n}= \bar{c}_n$, which reflects the fact that the phase-modulation is a real signal. 
Here, the number of data points $N$ is assumed to be even. If this number is odd, the term $N/2$ should be replaced with $(N-1)/2$.  
Due to the linearity of the HT, the reconstructed phase-modulation can be written as
\begin{eqnarray}
    u^{\rm H}[k] \approx 
    \sum_{n=0}^{N/2} f \left( v_n[k]\right).     \label{eq:uH-f-v}
\end{eqnarray}
We can further calculate Eq.(\ref{eq:uH-f-v}) for each term $f\left( v_n[k]\right)$ by dividing three cases based on a frequency index $m=\hat{\omega}N \Delta t / 2\pi$ that corresponds to the effective frequency $\hat{\omega}$. 

\begin{description}
\item[Case 1] Low frequency modulation: $ 0\le n \le m-1 $. 
    Using Bedrosian's theorem (Eq.~\ref{eq:Bedrosian_Th}) and Eq.(\ref{eq:HT-prop-sin}), we have 
    \begin{eqnarray}
		\dhilbert{v_n[k] \sin (\hat{\omega} k \Delta t)} = - v_n[k]\cos( \hat{\omega} k \Delta t ).  \label{eq:iH-case1}
	\end{eqnarray} 
    Substituting Eq.(\ref{eq:iH-case1}) into Eq.(\ref{eq:def-f}), we obtain 
    \begin{eqnarray}
		f \left( v_n[k]\right)= v_n[k]. \label{eq:uHk-case1}
	\end{eqnarray}
\item[Case 2] Middle frequency modulation: $ n=m$. 
   Using Bedrosian's theorem,  we have
   \begin{eqnarray}
       \dhilbert{v_n[k] \sin (\hat{\omega} k \Delta t)} = -\frac{1}{2} \left( c_n e^{2ik\omega_m } + \bar{c}_n e^{-2ik\omega_m }\right), \label{eq:iH-case2}
   \end{eqnarray}
   where the observation is assumed to be sufficiently long: $N > 4m$.
   Substituting Eq.(\ref{eq:iH-case2}) into Eq.(\ref{eq:def-f}), we obtain    
    \begin{eqnarray}
        f \left( v_n[k]\right)= 
          \left( \frac{3}{4}c_n -\frac{1}{4} \bar{c}_n \right) e^{i k \omega_n } - 
          \left( \frac{1}{4}c_n -\frac{3}{4}\bar{c}_n \right)e^{ - i k \omega_n }. 
        \label{eq:uHk-case2}
    \end{eqnarray}
\item[Case 3]: High frequency modulation: $m+1\le n \le N/2$. 
    Using Bedrosian's theorem, we have
   	\begin{eqnarray}
   	    \dhilbert{v_n[k] \sin (\hat{\omega} k \Delta t)} =
   	    \begin{cases}
   	        i \left( - c_n e^{ik\omega_{n} }  + \bar{c}_n e^{-ik\omega_{n}} \right) \sin (\hat{\omega} k \Delta t ) & {\rm for} \quad m+1 \le n < N/2, \\
   	        - i c_n e^{ik\omega_{n} }  \sin (\hat{\omega} k \Delta t )  & {\rm for} \quad n=N/2.
   	    \end{cases}\label{eq:iH-case3}
   	\end{eqnarray}
    Substituting Eq.(\ref{eq:iH-case3}) into Eq.(\ref{eq:def-f}), we obtain 
    \begin{eqnarray}
        f \left( v_n[k]\right)
        = \frac{1}{2}c_n e^{ik\omega_n} + \frac{1}{2}\bar{c}_n e^{-ik\omega_n} - \frac{1}{2} c_n e^{i k \omega_{n - 2m}} - \frac{1}{2} \bar{c}_n e^{-i k \omega_{n-2m} }.   \label{eq:uHk-case3}
    \end{eqnarray}
\end{description} 
Substituting Eqs.(\ref{eq:uHk-case1}), (\ref{eq:uHk-case2}), and (\ref{eq:uHk-case3}) into Eq.(\ref{eq:uH-f-v}), we obtain the Fourier series of the phase-modulation reconstructed by the conventional HT method 
\begin{eqnarray}
	u^{\rm H}[k]
	&\approx& c_{0}  - \frac{1}{2}\bar{c}_{2m} - \frac{1}{2}c_{2m}
	    + \left\{ \sum_{n=1}^{m-1}\left( c_{n} - \frac{1}{2} \bar{c}_{2m-n} - \frac{1}{2} c_{n+2m} \right) e^{ik\omega_n}
	    + \left(  \frac{3}{4} c_{m} -\frac{1}{4}\bar{c}_{m} - \frac{1}{2}c_{3m} \right)e^{ik \omega_m  } \right. \nonumber\\
	&& + \left. \sum_{ n= m+1 }^{N/2-2m}\left( \frac{1}{2}c_{n} - \frac{1}{2}c_{n+2m}\right)e^{ik\omega_n} 
        + \sum_{n=N/2-2m +1}^{N/2}\frac{1}{2}c_{n}e^{ik\omega_n} + {\rm c.c.} \right\},   
        \label{eq:uHk-results}
\end{eqnarray} 
where c.c denotes the complex conjugate of the terms in the curly brackets.
From Eq.(\ref{eq:uHk-results}), we finally obtain Eq.(\ref{eq:Ch}) by neglecting the higher order terms $O(\epsilon^2)$, which are expected to be small when the phase-modulation is weak.
%
Note that we can also obtain a similar formula between $c_n$ and $c^H_n$ (Eq. \ref{eq:Ch}) even when the amplitude $A(t)$ is changes slowly, i.e., the amplitude is a low-pass signal whose spectrum does not overlap with the spectra of $\cos\phi(t)$.

\subsection{Phase reconstruction based on the discrete Hilbert transform}
%
Here, we describe the conventional HT method for reconstructing the phase $\phi^{\rm H}[k]$ from a discrete oscillatory signal $x[k]$.
First, we pre-process the signal in order to mitigate the Gibbs phenomenon: we detect the first and the last peak points of the signal, and delete all of the data points prior to the first peak or after the last peak. 
Next, we calculate the Fourier transform of $x[k]$ to obtain $X_n$. 
Then, we calculate the discrete HT $\dhilbert{x[k]}$ according to Eq.(\ref{eq:def-dHT1}).
Finally, we reconstruct the phase by calculating the argument of the analytic signal $\phi^{\rm H}[k] := \arg \left( x[k] + i\dhilbert{x[k]} \right)$.

\section{Data availability}
The datasets used during this study are available from the corresponding author on reasonable request.

\section{Code availability}
The code of the extended Hilbert transform is available from the corresponding author on reasonable request.

\bibliography{ref_pe.bib}

\begin{thebibliography}{45}%
\makeatletter
\providecommand \@ifxundefined [1]{%
 \@ifx{#1\undefined}
}%
\providecommand \@ifnum [1]{%
 \ifnum #1\expandafter \@firstoftwo
 \else \expandafter \@secondoftwo
 \fi
}%
\providecommand \@ifx [1]{%
 \ifx #1\expandafter \@firstoftwo
 \else \expandafter \@secondoftwo
 \fi
}%
\providecommand \natexlab [1]{#1}%
\providecommand \enquote  [1]{``#1''}%
\providecommand \bibnamefont  [1]{#1}%
\providecommand \bibfnamefont [1]{#1}%
\providecommand \citenamefont [1]{#1}%
\providecommand \href@noop [0]{\@secondoftwo}%
\providecommand \href [0]{\begingroup \@sanitize@url \@href}%
\providecommand \@href[1]{\@@startlink{#1}\@@href}%
\providecommand \@@href[1]{\endgroup#1\@@endlink}%
\providecommand \@sanitize@url [0]{\catcode `\\12\catcode `\$12\catcode
  `\&12\catcode `\#12\catcode `\^12\catcode `\_12\catcode `\%12\relax}%
\providecommand \@@startlink[1]{}%
\providecommand \@@endlink[0]{}%
\providecommand \url  [0]{\begingroup\@sanitize@url \@url }%
\providecommand \@url [1]{\endgroup\@href {#1}{\urlprefix }}%
\providecommand \urlprefix  [0]{URL }%
\providecommand \Eprint [0]{\href }%
\providecommand \doibase [0]{https://doi.org/}%
\providecommand \selectlanguage [0]{\@gobble}%
\providecommand \bibinfo  [0]{\@secondoftwo}%
\providecommand \bibfield  [0]{\@secondoftwo}%
\providecommand \translation [1]{[#1]}%
\providecommand \BibitemOpen [0]{}%
\providecommand \bibitemStop [0]{}%
\providecommand \bibitemNoStop [0]{.\EOS\space}%
\providecommand \EOS [0]{\spacefactor3000\relax}%
\providecommand \BibitemShut  [1]{\csname bibitem#1\endcsname}%
\let\auto@bib@innerbib\@empty
\bibitem [{\citenamefont {Buzsaki}\ and\ \citenamefont
  {Draguhn}(2004)}]{buzsaki2004neuronal}%
  \BibitemOpen
  \bibfield  {author} {\bibinfo {author} {\bibfnamefont {G.}~\bibnamefont
  {Buzsaki}}\ and\ \bibinfo {author} {\bibfnamefont {A.}~\bibnamefont
  {Draguhn}},\ }\bibfield  {title} {\bibinfo {title} {Neuronal oscillations in
  cortical networks},\ }\href@noop {} {\bibfield  {journal} {\bibinfo
  {journal} {Science}\ }\textbf {\bibinfo {volume} {304}},\ \bibinfo {pages}
  {1926} (\bibinfo {year} {2004})}\BibitemShut {NoStop}%
\bibitem [{\citenamefont {Wang}(2010)}]{wang2010}%
  \BibitemOpen
  \bibfield  {author} {\bibinfo {author} {\bibfnamefont {X.-J.}\ \bibnamefont
  {Wang}},\ }\bibfield  {title} {\bibinfo {title} {Neurophysiological and
  computational principles of cortical rhythms in cognition},\ }\href@noop {}
  {\bibfield  {journal} {\bibinfo  {journal} {Physiological reviews}\ }\textbf
  {\bibinfo {volume} {90}},\ \bibinfo {pages} {1195} (\bibinfo {year}
  {2010})}\BibitemShut {NoStop}%
\bibitem [{\citenamefont {Sch{\"a}fer}\ \emph {et~al.}(1998)\citenamefont
  {Sch{\"a}fer}, \citenamefont {Rosenblum}, \citenamefont {Kurths},\ and\
  \citenamefont {Abel}}]{schafer1998}%
  \BibitemOpen
  \bibfield  {author} {\bibinfo {author} {\bibfnamefont {C.}~\bibnamefont
  {Sch{\"a}fer}}, \bibinfo {author} {\bibfnamefont {M.~G.}\ \bibnamefont
  {Rosenblum}}, \bibinfo {author} {\bibfnamefont {J.}~\bibnamefont {Kurths}},\
  and\ \bibinfo {author} {\bibfnamefont {H.-H.}\ \bibnamefont {Abel}},\
  }\bibfield  {title} {\bibinfo {title} {Heartbeat synchronized with
  ventilation},\ }\href@noop {} {\bibfield  {journal} {\bibinfo  {journal}
  {Nature}\ }\textbf {\bibinfo {volume} {392}},\ \bibinfo {pages} {239}
  (\bibinfo {year} {1998})}\BibitemShut {NoStop}%
\bibitem [{\citenamefont {Lotri{\v{c}}}\ and\ \citenamefont
  {Stefanovska}(2000)}]{lotrivc2000}%
  \BibitemOpen
  \bibfield  {author} {\bibinfo {author} {\bibfnamefont {M.~B.}\ \bibnamefont
  {Lotri{\v{c}}}}\ and\ \bibinfo {author} {\bibfnamefont {A.}~\bibnamefont
  {Stefanovska}},\ }\bibfield  {title} {\bibinfo {title} {Synchronization and
  modulation in the human cardiorespiratory system},\ }\href@noop {} {\bibfield
   {journal} {\bibinfo  {journal} {Physica A: Statistical Mechanics and its
  Applications}\ }\textbf {\bibinfo {volume} {283}},\ \bibinfo {pages} {451}
  (\bibinfo {year} {2000})}\BibitemShut {NoStop}%
\bibitem [{\citenamefont {Kralemann}\ \emph {et~al.}(2013)\citenamefont
  {Kralemann}, \citenamefont {Fr{\"u}hwirth}, \citenamefont {Pikovsky},
  \citenamefont {Rosenblum}, \citenamefont {Kenner}, \citenamefont {Schaefer},\
  and\ \citenamefont {Moser}}]{kralemann2013vivo}%
  \BibitemOpen
  \bibfield  {author} {\bibinfo {author} {\bibfnamefont {B.}~\bibnamefont
  {Kralemann}}, \bibinfo {author} {\bibfnamefont {M.}~\bibnamefont
  {Fr{\"u}hwirth}}, \bibinfo {author} {\bibfnamefont {A.}~\bibnamefont
  {Pikovsky}}, \bibinfo {author} {\bibfnamefont {M.}~\bibnamefont {Rosenblum}},
  \bibinfo {author} {\bibfnamefont {T.}~\bibnamefont {Kenner}}, \bibinfo
  {author} {\bibfnamefont {J.}~\bibnamefont {Schaefer}},\ and\ \bibinfo
  {author} {\bibfnamefont {M.}~\bibnamefont {Moser}},\ }\bibfield  {title}
  {\bibinfo {title} {In vivo cardiac phase response curve elucidates human
  respiratory heart rate variability},\ }\href@noop {} {\bibfield  {journal}
  {\bibinfo  {journal} {Nature communications}\ }\textbf {\bibinfo {volume}
  {4}},\ \bibinfo {pages} {1} (\bibinfo {year} {2013})}\BibitemShut {NoStop}%
\bibitem [{\citenamefont {Fukuda}\ \emph {et~al.}(2013)\citenamefont {Fukuda},
  \citenamefont {Murase},\ and\ \citenamefont
  {Tokuda}}]{fukuda2013controlling}%
  \BibitemOpen
  \bibfield  {author} {\bibinfo {author} {\bibfnamefont {H.}~\bibnamefont
  {Fukuda}}, \bibinfo {author} {\bibfnamefont {H.}~\bibnamefont {Murase}},\
  and\ \bibinfo {author} {\bibfnamefont {I.~T.}\ \bibnamefont {Tokuda}},\
  }\bibfield  {title} {\bibinfo {title} {Controlling circadian rhythms by
  dark-pulse perturbations in arabidopsis thaliana},\ }\href@noop {} {\bibfield
   {journal} {\bibinfo  {journal} {Scientific Reports}\ }\textbf {\bibinfo
  {volume} {3}},\ \bibinfo {pages} {1} (\bibinfo {year} {2013})}\BibitemShut
  {NoStop}%
\bibitem [{\citenamefont {Yamaguchi}\ \emph {et~al.}(2013)\citenamefont
  {Yamaguchi}, \citenamefont {Suzuki}, \citenamefont {Mizoro}, \citenamefont
  {Kori}, \citenamefont {Okada}, \citenamefont {Chen}, \citenamefont {Fustin},
  \citenamefont {Yamazaki}, \citenamefont {Mizuguchi}, \citenamefont {Zhang}
  \emph {et~al.}}]{yamaguchi2013}%
  \BibitemOpen
  \bibfield  {author} {\bibinfo {author} {\bibfnamefont {Y.}~\bibnamefont
  {Yamaguchi}}, \bibinfo {author} {\bibfnamefont {T.}~\bibnamefont {Suzuki}},
  \bibinfo {author} {\bibfnamefont {Y.}~\bibnamefont {Mizoro}}, \bibinfo
  {author} {\bibfnamefont {H.}~\bibnamefont {Kori}}, \bibinfo {author}
  {\bibfnamefont {K.}~\bibnamefont {Okada}}, \bibinfo {author} {\bibfnamefont
  {Y.}~\bibnamefont {Chen}}, \bibinfo {author} {\bibfnamefont {J.-M.}\
  \bibnamefont {Fustin}}, \bibinfo {author} {\bibfnamefont {F.}~\bibnamefont
  {Yamazaki}}, \bibinfo {author} {\bibfnamefont {N.}~\bibnamefont {Mizuguchi}},
  \bibinfo {author} {\bibfnamefont {J.}~\bibnamefont {Zhang}}, \emph {et~al.},\
  }\bibfield  {title} {\bibinfo {title} {Mice genetically deficient in
  vasopressin v1a and v1b receptors are resistant to jet lag},\ }\href@noop {}
  {\bibfield  {journal} {\bibinfo  {journal} {Science}\ }\textbf {\bibinfo
  {volume} {342}},\ \bibinfo {pages} {85} (\bibinfo {year} {2013})}\BibitemShut
  {NoStop}%
\bibitem [{\citenamefont {Yoshioka-Kobayashi}\ \emph
  {et~al.}(2020)\citenamefont {Yoshioka-Kobayashi}, \citenamefont {Matsumiya},
  \citenamefont {Niino}, \citenamefont {Isomura}, \citenamefont {Kori},
  \citenamefont {Miyawaki},\ and\ \citenamefont
  {Kageyama}}]{yoshioka2020coupling}%
  \BibitemOpen
  \bibfield  {author} {\bibinfo {author} {\bibfnamefont {K.}~\bibnamefont
  {Yoshioka-Kobayashi}}, \bibinfo {author} {\bibfnamefont {M.}~\bibnamefont
  {Matsumiya}}, \bibinfo {author} {\bibfnamefont {Y.}~\bibnamefont {Niino}},
  \bibinfo {author} {\bibfnamefont {A.}~\bibnamefont {Isomura}}, \bibinfo
  {author} {\bibfnamefont {H.}~\bibnamefont {Kori}}, \bibinfo {author}
  {\bibfnamefont {A.}~\bibnamefont {Miyawaki}},\ and\ \bibinfo {author}
  {\bibfnamefont {R.}~\bibnamefont {Kageyama}},\ }\bibfield  {title} {\bibinfo
  {title} {Coupling delay controls synchronized oscillation in the segmentation
  clock},\ }\href@noop {} {\bibfield  {journal} {\bibinfo  {journal} {Nature}\
  }\textbf {\bibinfo {volume} {580}},\ \bibinfo {pages} {119} (\bibinfo {year}
  {2020})}\BibitemShut {NoStop}%
\bibitem [{\citenamefont {Collins}\ and\ \citenamefont
  {Stewart}(1993)}]{collins1993}%
  \BibitemOpen
  \bibfield  {author} {\bibinfo {author} {\bibfnamefont {J.~J.}\ \bibnamefont
  {Collins}}\ and\ \bibinfo {author} {\bibfnamefont {I.~N.}\ \bibnamefont
  {Stewart}},\ }\bibfield  {title} {\bibinfo {title} {Coupled nonlinear
  oscillators and the symmetries of animal gaits},\ }\href@noop {} {\bibfield
  {journal} {\bibinfo  {journal} {Journal of Nonlinear Science}\ }\textbf
  {\bibinfo {volume} {3}},\ \bibinfo {pages} {349} (\bibinfo {year}
  {1993})}\BibitemShut {NoStop}%
\bibitem [{\citenamefont {Borgius}\ \emph {et~al.}(2014)\citenamefont
  {Borgius}, \citenamefont {Nishimaru}, \citenamefont {Caldeira}, \citenamefont
  {Kunugise}, \citenamefont {L{\"o}w}, \citenamefont {Reig}, \citenamefont
  {Itohara}, \citenamefont {Iwasato},\ and\ \citenamefont
  {Kiehn}}]{borgius2014spinal}%
  \BibitemOpen
  \bibfield  {author} {\bibinfo {author} {\bibfnamefont {L.}~\bibnamefont
  {Borgius}}, \bibinfo {author} {\bibfnamefont {H.}~\bibnamefont {Nishimaru}},
  \bibinfo {author} {\bibfnamefont {V.}~\bibnamefont {Caldeira}}, \bibinfo
  {author} {\bibfnamefont {Y.}~\bibnamefont {Kunugise}}, \bibinfo {author}
  {\bibfnamefont {P.}~\bibnamefont {L{\"o}w}}, \bibinfo {author} {\bibfnamefont
  {R.}~\bibnamefont {Reig}}, \bibinfo {author} {\bibfnamefont {S.}~\bibnamefont
  {Itohara}}, \bibinfo {author} {\bibfnamefont {T.}~\bibnamefont {Iwasato}},\
  and\ \bibinfo {author} {\bibfnamefont {O.}~\bibnamefont {Kiehn}},\ }\bibfield
   {title} {\bibinfo {title} {Spinal glutamatergic neurons defined by epha4
  signaling are essential components of normal locomotor circuits},\
  }\href@noop {} {\bibfield  {journal} {\bibinfo  {journal} {Journal of
  Neuroscience}\ }\textbf {\bibinfo {volume} {34}},\ \bibinfo {pages} {3841}
  (\bibinfo {year} {2014})}\BibitemShut {NoStop}%
\bibitem [{\citenamefont {Kobayashi}\ \emph {et~al.}(2016)\citenamefont
  {Kobayashi}, \citenamefont {Nishimaru},\ and\ \citenamefont
  {Nishijo}}]{kobayashi2016}%
  \BibitemOpen
  \bibfield  {author} {\bibinfo {author} {\bibfnamefont {R.}~\bibnamefont
  {Kobayashi}}, \bibinfo {author} {\bibfnamefont {H.}~\bibnamefont
  {Nishimaru}},\ and\ \bibinfo {author} {\bibfnamefont {H.}~\bibnamefont
  {Nishijo}},\ }\bibfield  {title} {\bibinfo {title} {Estimation of excitatory
  and inhibitory synaptic conductance variations in motoneurons during
  locomotor-like rhythmic activity},\ }\href@noop {} {\bibfield  {journal}
  {\bibinfo  {journal} {Neuroscience}\ }\textbf {\bibinfo {volume} {335}},\
  \bibinfo {pages} {72} (\bibinfo {year} {2016})}\BibitemShut {NoStop}%
\bibitem [{\citenamefont {Winfree}(1980)}]{winfree1980}%
  \BibitemOpen
  \bibfield  {author} {\bibinfo {author} {\bibfnamefont {A.~T.}\ \bibnamefont
  {Winfree}},\ }\href@noop {} {\emph {\bibinfo {title} {The geometry of
  biological time}}},\ Vol.~\bibinfo {volume} {2}\ (\bibinfo  {publisher}
  {Springer},\ \bibinfo {year} {1980})\BibitemShut {NoStop}%
\bibitem [{\citenamefont {Kuramoto}(1984)}]{kuramoto1984chemical}%
  \BibitemOpen
  \bibfield  {author} {\bibinfo {author} {\bibfnamefont {Y.}~\bibnamefont
  {Kuramoto}},\ }\href@noop {} {\emph {\bibinfo {title} {Chemical Oscillations,
  Waves and Turbulence}}}\ (\bibinfo  {publisher} {Springer, Berlin},\ \bibinfo
  {year} {1984})\BibitemShut {NoStop}%
\bibitem [{\citenamefont {Ashwin}\ \emph {et~al.}(2016)\citenamefont {Ashwin},
  \citenamefont {Coombes},\ and\ \citenamefont {Nicks}}]{ashwin2016}%
  \BibitemOpen
  \bibfield  {author} {\bibinfo {author} {\bibfnamefont {P.}~\bibnamefont
  {Ashwin}}, \bibinfo {author} {\bibfnamefont {S.}~\bibnamefont {Coombes}},\
  and\ \bibinfo {author} {\bibfnamefont {R.}~\bibnamefont {Nicks}},\ }\bibfield
   {title} {\bibinfo {title} {Mathematical frameworks for oscillatory network
  dynamics in neuroscience},\ }\href@noop {} {\bibfield  {journal} {\bibinfo
  {journal} {The Journal of Mathematical Neuroscience}\ }\textbf {\bibinfo
  {volume} {6}},\ \bibinfo {pages} {1} (\bibinfo {year} {2016})}\BibitemShut
  {NoStop}%
\bibitem [{\citenamefont {Nakao}(2016)}]{nakao2016}%
  \BibitemOpen
  \bibfield  {author} {\bibinfo {author} {\bibfnamefont {H.}~\bibnamefont
  {Nakao}},\ }\bibfield  {title} {\bibinfo {title} {Phase reduction approach to
  synchronisation of nonlinear oscillators},\ }\href@noop {} {\bibfield
  {journal} {\bibinfo  {journal} {Contemporary Physics}\ }\textbf {\bibinfo
  {volume} {57}},\ \bibinfo {pages} {188} (\bibinfo {year} {2016})}\BibitemShut
  {NoStop}%
\bibitem [{\citenamefont {Pikovsky}\ \emph {et~al.}(2003)\citenamefont
  {Pikovsky}, \citenamefont {Kurths}, \citenamefont {Rosenblum},\ and\
  \citenamefont {Kurths}}]{pikovsky2003synchronization}%
  \BibitemOpen
  \bibfield  {author} {\bibinfo {author} {\bibfnamefont {A.}~\bibnamefont
  {Pikovsky}}, \bibinfo {author} {\bibfnamefont {J.}~\bibnamefont {Kurths}},
  \bibinfo {author} {\bibfnamefont {M.}~\bibnamefont {Rosenblum}},\ and\
  \bibinfo {author} {\bibfnamefont {J.}~\bibnamefont {Kurths}},\ }\href@noop {}
  {\emph {\bibinfo {title} {Synchronization: a universal concept in nonlinear
  sciences}}},\ \bibinfo {number} {12}\ (\bibinfo  {publisher} {Cambridge
  university press},\ \bibinfo {year} {2003})\BibitemShut {NoStop}%
\bibitem [{\citenamefont {Gal{\'a}n}\ \emph {et~al.}(2005)\citenamefont
  {Gal{\'a}n}, \citenamefont {Ermentrout},\ and\ \citenamefont
  {Urban}}]{galan2005efficient}%
  \BibitemOpen
  \bibfield  {author} {\bibinfo {author} {\bibfnamefont {R.~F.}\ \bibnamefont
  {Gal{\'a}n}}, \bibinfo {author} {\bibfnamefont {G.~B.}\ \bibnamefont
  {Ermentrout}},\ and\ \bibinfo {author} {\bibfnamefont {N.~N.}\ \bibnamefont
  {Urban}},\ }\bibfield  {title} {\bibinfo {title} {Efficient estimation of
  phase-resetting curves in real neurons and its significance for
  neural-network modeling},\ }\href@noop {} {\bibfield  {journal} {\bibinfo
  {journal} {Physical Review Letters}\ }\textbf {\bibinfo {volume} {94}},\
  \bibinfo {pages} {158101} (\bibinfo {year} {2005})}\BibitemShut {NoStop}%
\bibitem [{\citenamefont {Ota}\ \emph {et~al.}(2009)\citenamefont {Ota},
  \citenamefont {Nomura},\ and\ \citenamefont {Aoyagi}}]{ota2009weighted}%
  \BibitemOpen
  \bibfield  {author} {\bibinfo {author} {\bibfnamefont {K.}~\bibnamefont
  {Ota}}, \bibinfo {author} {\bibfnamefont {M.}~\bibnamefont {Nomura}},\ and\
  \bibinfo {author} {\bibfnamefont {T.}~\bibnamefont {Aoyagi}},\ }\bibfield
  {title} {\bibinfo {title} {Weighted spike-triggered average of a fluctuating
  stimulus yielding the phase response curve},\ }\href@noop {} {\bibfield
  {journal} {\bibinfo  {journal} {Physical Review Letters}\ }\textbf {\bibinfo
  {volume} {103}},\ \bibinfo {pages} {024101} (\bibinfo {year}
  {2009})}\BibitemShut {NoStop}%
\bibitem [{\citenamefont {Nakae}\ \emph {et~al.}(2010)\citenamefont {Nakae},
  \citenamefont {Iba}, \citenamefont {Tsubo}, \citenamefont {Fukai},\ and\
  \citenamefont {Aoyagi}}]{nakae2010bayesian}%
  \BibitemOpen
  \bibfield  {author} {\bibinfo {author} {\bibfnamefont {K.}~\bibnamefont
  {Nakae}}, \bibinfo {author} {\bibfnamefont {Y.}~\bibnamefont {Iba}}, \bibinfo
  {author} {\bibfnamefont {Y.}~\bibnamefont {Tsubo}}, \bibinfo {author}
  {\bibfnamefont {T.}~\bibnamefont {Fukai}},\ and\ \bibinfo {author}
  {\bibfnamefont {T.}~\bibnamefont {Aoyagi}},\ }\bibfield  {title} {\bibinfo
  {title} {Bayesian estimation of phase response curves},\ }\href@noop {}
  {\bibfield  {journal} {\bibinfo  {journal} {Neural Networks}\ }\textbf
  {\bibinfo {volume} {23}},\ \bibinfo {pages} {752} (\bibinfo {year}
  {2010})}\BibitemShut {NoStop}%
\bibitem [{\citenamefont {Cestnik}\ and\ \citenamefont
  {Rosenblum}(2018)}]{cestnik2018inferring}%
  \BibitemOpen
  \bibfield  {author} {\bibinfo {author} {\bibfnamefont {R.}~\bibnamefont
  {Cestnik}}\ and\ \bibinfo {author} {\bibfnamefont {M.}~\bibnamefont
  {Rosenblum}},\ }\bibfield  {title} {\bibinfo {title} {Inferring the phase
  response curve from observation of a continuously perturbed oscillator},\
  }\href@noop {} {\bibfield  {journal} {\bibinfo  {journal} {Scientific
  Reports}\ }\textbf {\bibinfo {volume} {8}},\ \bibinfo {pages} {1} (\bibinfo
  {year} {2018})}\BibitemShut {NoStop}%
\bibitem [{\citenamefont {Namura}\ \emph {et~al.}(2022)\citenamefont {Namura},
  \citenamefont {Takata}, \citenamefont {Yamaguchi}, \citenamefont
  {Kobayashi},\ and\ \citenamefont {Nakao}}]{namura2022estimating}%
  \BibitemOpen
  \bibfield  {author} {\bibinfo {author} {\bibfnamefont {N.}~\bibnamefont
  {Namura}}, \bibinfo {author} {\bibfnamefont {S.}~\bibnamefont {Takata}},
  \bibinfo {author} {\bibfnamefont {K.}~\bibnamefont {Yamaguchi}}, \bibinfo
  {author} {\bibfnamefont {R.}~\bibnamefont {Kobayashi}},\ and\ \bibinfo
  {author} {\bibfnamefont {H.}~\bibnamefont {Nakao}},\ }\bibfield  {title}
  {\bibinfo {title} {Estimating asymptotic phase and amplitude functions of
  limit-cycle oscillators from time series data},\ }\href@noop {} {\bibfield
  {journal} {\bibinfo  {journal} {Physical Review E}\ }\textbf {\bibinfo
  {volume} {106}},\ \bibinfo {pages} {014204} (\bibinfo {year}
  {2022})}\BibitemShut {NoStop}%
\bibitem [{\citenamefont {Rosenblum}\ and\ \citenamefont
  {Pikovsky}(2001)}]{rosenblum2001detecting}%
  \BibitemOpen
  \bibfield  {author} {\bibinfo {author} {\bibfnamefont {M.~G.}\ \bibnamefont
  {Rosenblum}}\ and\ \bibinfo {author} {\bibfnamefont {A.~S.}\ \bibnamefont
  {Pikovsky}},\ }\bibfield  {title} {\bibinfo {title} {Detecting direction of
  coupling in interacting oscillators},\ }\href@noop {} {\bibfield  {journal}
  {\bibinfo  {journal} {Physical Review E}\ }\textbf {\bibinfo {volume} {64}},\
  \bibinfo {pages} {045202} (\bibinfo {year} {2001})}\BibitemShut {NoStop}%
\bibitem [{\citenamefont {Tokuda}\ \emph {et~al.}(2007)\citenamefont {Tokuda},
  \citenamefont {Jain}, \citenamefont {Kiss},\ and\ \citenamefont
  {Hudson}}]{tokuda2007inferring}%
  \BibitemOpen
  \bibfield  {author} {\bibinfo {author} {\bibfnamefont {I.~T.}\ \bibnamefont
  {Tokuda}}, \bibinfo {author} {\bibfnamefont {S.}~\bibnamefont {Jain}},
  \bibinfo {author} {\bibfnamefont {I.~Z.}\ \bibnamefont {Kiss}},\ and\
  \bibinfo {author} {\bibfnamefont {J.~L.}\ \bibnamefont {Hudson}},\ }\bibfield
   {title} {\bibinfo {title} {Inferring phase equations from multivariate time
  series},\ }\href@noop {} {\bibfield  {journal} {\bibinfo  {journal} {Physical
  Review Letters}\ }\textbf {\bibinfo {volume} {99}},\ \bibinfo {pages}
  {064101} (\bibinfo {year} {2007})}\BibitemShut {NoStop}%
\bibitem [{\citenamefont {Kralemann}\ \emph {et~al.}(2008)\citenamefont
  {Kralemann}, \citenamefont {Cimponeriu}, \citenamefont {Rosenblum},
  \citenamefont {Pikovsky},\ and\ \citenamefont {Mrowka}}]{kralemann2008phase}%
  \BibitemOpen
  \bibfield  {author} {\bibinfo {author} {\bibfnamefont {B.}~\bibnamefont
  {Kralemann}}, \bibinfo {author} {\bibfnamefont {L.}~\bibnamefont
  {Cimponeriu}}, \bibinfo {author} {\bibfnamefont {M.}~\bibnamefont
  {Rosenblum}}, \bibinfo {author} {\bibfnamefont {A.}~\bibnamefont
  {Pikovsky}},\ and\ \bibinfo {author} {\bibfnamefont {R.}~\bibnamefont
  {Mrowka}},\ }\bibfield  {title} {\bibinfo {title} {Phase dynamics of coupled
  oscillators reconstructed from data},\ }\href@noop {} {\bibfield  {journal}
  {\bibinfo  {journal} {Physical Review E}\ }\textbf {\bibinfo {volume} {77}},\
  \bibinfo {pages} {066205} (\bibinfo {year} {2008})}\BibitemShut {NoStop}%
\bibitem [{\citenamefont {Ren}\ \emph {et~al.}(2010)\citenamefont {Ren},
  \citenamefont {Wang}, \citenamefont {Li},\ and\ \citenamefont
  {Lai}}]{ren2010noise}%
  \BibitemOpen
  \bibfield  {author} {\bibinfo {author} {\bibfnamefont {J.}~\bibnamefont
  {Ren}}, \bibinfo {author} {\bibfnamefont {W.-X.}\ \bibnamefont {Wang}},
  \bibinfo {author} {\bibfnamefont {B.}~\bibnamefont {Li}},\ and\ \bibinfo
  {author} {\bibfnamefont {Y.-C.}\ \bibnamefont {Lai}},\ }\bibfield  {title}
  {\bibinfo {title} {Noise bridges dynamical correlation and topology in
  coupled oscillator networks},\ }\href@noop {} {\bibfield  {journal} {\bibinfo
   {journal} {Physical review letters}\ }\textbf {\bibinfo {volume} {104}},\
  \bibinfo {pages} {058701} (\bibinfo {year} {2010})}\BibitemShut {NoStop}%
\bibitem [{\citenamefont {Levnaji{\'c}}\ and\ \citenamefont
  {Pikovsky}(2011)}]{levnajic2011network}%
  \BibitemOpen
  \bibfield  {author} {\bibinfo {author} {\bibfnamefont {Z.}~\bibnamefont
  {Levnaji{\'c}}}\ and\ \bibinfo {author} {\bibfnamefont {A.}~\bibnamefont
  {Pikovsky}},\ }\bibfield  {title} {\bibinfo {title} {Network reconstruction
  from random phase resetting},\ }\href@noop {} {\bibfield  {journal} {\bibinfo
   {journal} {Physical Review Letters}\ }\textbf {\bibinfo {volume} {107}},\
  \bibinfo {pages} {034101} (\bibinfo {year} {2011})}\BibitemShut {NoStop}%
\bibitem [{\citenamefont {Stankovski}\ \emph {et~al.}(2012)\citenamefont
  {Stankovski}, \citenamefont {Duggento}, \citenamefont {McClintock},\ and\
  \citenamefont {Stefanovska}}]{stankovski2012inference}%
  \BibitemOpen
  \bibfield  {author} {\bibinfo {author} {\bibfnamefont {T.}~\bibnamefont
  {Stankovski}}, \bibinfo {author} {\bibfnamefont {A.}~\bibnamefont
  {Duggento}}, \bibinfo {author} {\bibfnamefont {P.~V.}\ \bibnamefont
  {McClintock}},\ and\ \bibinfo {author} {\bibfnamefont {A.}~\bibnamefont
  {Stefanovska}},\ }\bibfield  {title} {\bibinfo {title} {Inference of
  time-evolving coupled dynamical systems in the presence of noise},\
  }\href@noop {} {\bibfield  {journal} {\bibinfo  {journal} {Physical Review
  Letters}\ }\textbf {\bibinfo {volume} {109}},\ \bibinfo {pages} {024101}
  (\bibinfo {year} {2012})}\BibitemShut {NoStop}%
\bibitem [{\citenamefont {{\O}stergaard}\ \emph {et~al.}(2017)\citenamefont
  {{\O}stergaard}, \citenamefont {Rahbek},\ and\ \citenamefont
  {Ditlevsen}}]{ostergaard2017oscillating}%
  \BibitemOpen
  \bibfield  {author} {\bibinfo {author} {\bibfnamefont {J.}~\bibnamefont
  {{\O}stergaard}}, \bibinfo {author} {\bibfnamefont {A.}~\bibnamefont
  {Rahbek}},\ and\ \bibinfo {author} {\bibfnamefont {S.}~\bibnamefont
  {Ditlevsen}},\ }\bibfield  {title} {\bibinfo {title} {Oscillating systems
  with cointegrated phase processes},\ }\href@noop {} {\bibfield  {journal}
  {\bibinfo  {journal} {Journal of Mathematical Biology}\ }\textbf {\bibinfo
  {volume} {75}},\ \bibinfo {pages} {845} (\bibinfo {year} {2017})}\BibitemShut
  {NoStop}%
\bibitem [{\citenamefont {Onojima}\ \emph {et~al.}(2018)\citenamefont
  {Onojima}, \citenamefont {Goto}, \citenamefont {Mizuhara},\ and\
  \citenamefont {Aoyagi}}]{onojima2018dynamical}%
  \BibitemOpen
  \bibfield  {author} {\bibinfo {author} {\bibfnamefont {T.}~\bibnamefont
  {Onojima}}, \bibinfo {author} {\bibfnamefont {T.}~\bibnamefont {Goto}},
  \bibinfo {author} {\bibfnamefont {H.}~\bibnamefont {Mizuhara}},\ and\
  \bibinfo {author} {\bibfnamefont {T.}~\bibnamefont {Aoyagi}},\ }\bibfield
  {title} {\bibinfo {title} {A dynamical systems approach for estimating phase
  interactions between rhythms of different frequencies from experimental
  data},\ }\href@noop {} {\bibfield  {journal} {\bibinfo  {journal} {PLoS
  Computational Biology}\ }\textbf {\bibinfo {volume} {14}},\ \bibinfo {pages}
  {e1005928} (\bibinfo {year} {2018})}\BibitemShut {NoStop}%
\bibitem [{\citenamefont {Suzuki}\ \emph {et~al.}(2018)\citenamefont {Suzuki},
  \citenamefont {Aoyagi},\ and\ \citenamefont {Kitano}}]{suzuki2018bayesian}%
  \BibitemOpen
  \bibfield  {author} {\bibinfo {author} {\bibfnamefont {K.}~\bibnamefont
  {Suzuki}}, \bibinfo {author} {\bibfnamefont {T.}~\bibnamefont {Aoyagi}},\
  and\ \bibinfo {author} {\bibfnamefont {K.}~\bibnamefont {Kitano}},\
  }\bibfield  {title} {\bibinfo {title} {Bayesian estimation of phase dynamics
  based on partially sampled spikes generated by realistic model neurons},\
  }\href@noop {} {\bibfield  {journal} {\bibinfo  {journal} {Frontiers in
  Computational Neuroscience}\ }\textbf {\bibinfo {volume} {11}},\ \bibinfo
  {pages} {116} (\bibinfo {year} {2018})}\BibitemShut {NoStop}%
\bibitem [{\citenamefont {Stankovski}\ \emph {et~al.}(2017)\citenamefont
  {Stankovski}, \citenamefont {Pereira}, \citenamefont {McClintock},\ and\
  \citenamefont {Stefanovska}}]{stankovski2017coupling}%
  \BibitemOpen
  \bibfield  {author} {\bibinfo {author} {\bibfnamefont {T.}~\bibnamefont
  {Stankovski}}, \bibinfo {author} {\bibfnamefont {T.}~\bibnamefont {Pereira}},
  \bibinfo {author} {\bibfnamefont {P.~V.}\ \bibnamefont {McClintock}},\ and\
  \bibinfo {author} {\bibfnamefont {A.}~\bibnamefont {Stefanovska}},\
  }\bibfield  {title} {\bibinfo {title} {Coupling functions: universal insights
  into dynamical interaction mechanisms},\ }\href@noop {} {\bibfield  {journal}
  {\bibinfo  {journal} {Reviews of Modern Physics}\ }\textbf {\bibinfo {volume}
  {89}},\ \bibinfo {pages} {045001} (\bibinfo {year} {2017})}\BibitemShut
  {NoStop}%
\bibitem [{\citenamefont {Tokuda}\ \emph {et~al.}(2019)\citenamefont {Tokuda},
  \citenamefont {Levnajic},\ and\ \citenamefont
  {Ishimura}}]{tokuda2019practical}%
  \BibitemOpen
  \bibfield  {author} {\bibinfo {author} {\bibfnamefont {I.~T.}\ \bibnamefont
  {Tokuda}}, \bibinfo {author} {\bibfnamefont {Z.}~\bibnamefont {Levnajic}},\
  and\ \bibinfo {author} {\bibfnamefont {K.}~\bibnamefont {Ishimura}},\
  }\bibfield  {title} {\bibinfo {title} {A practical method for estimating
  coupling functions in complex dynamical systems},\ }\href@noop {} {\bibfield
  {journal} {\bibinfo  {journal} {Philosophical Transactions of the Royal
  Society A}\ }\textbf {\bibinfo {volume} {377}},\ \bibinfo {pages} {20190015}
  (\bibinfo {year} {2019})}\BibitemShut {NoStop}%
\bibitem [{\citenamefont {Gengel}\ and\ \citenamefont
  {Pikovsky}(2019)}]{gengel2019phase}%
  \BibitemOpen
  \bibfield  {author} {\bibinfo {author} {\bibfnamefont {E.}~\bibnamefont
  {Gengel}}\ and\ \bibinfo {author} {\bibfnamefont {A.}~\bibnamefont
  {Pikovsky}},\ }\bibfield  {title} {\bibinfo {title} {Phase demodulation with
  iterative Hilbert transform embeddings},\ }\href@noop {} {\bibfield
  {journal} {\bibinfo  {journal} {Signal Processing}\ }\textbf {\bibinfo
  {volume} {165}},\ \bibinfo {pages} {115} (\bibinfo {year}
  {2019})}\BibitemShut {NoStop}%
\bibitem [{\citenamefont {Gabor}(1946)}]{gabor1946theory}%
  \BibitemOpen
  \bibfield  {author} {\bibinfo {author} {\bibfnamefont {D.}~\bibnamefont
  {Gabor}},\ }\bibfield  {title} {\bibinfo {title} {Theory of communication.
  part 3: Frequency compression and expansion},\ }\href@noop {} {\bibfield
  {journal} {\bibinfo  {journal} {Journal of the Institution of Electrical
  Engineers-part III: radio and communication engineering}\ }\textbf {\bibinfo
  {volume} {93}},\ \bibinfo {pages} {445} (\bibinfo {year} {1946})}\BibitemShut
  {NoStop}%
\bibitem [{\citenamefont {King}(2009)}]{king2009hilbert-vol1}%
  \BibitemOpen
  \bibfield  {author} {\bibinfo {author} {\bibfnamefont {F.~W.}\ \bibnamefont
  {King}},\ }\href@noop {} {\emph {\bibinfo {title} {Hilbert Transforms}}},\
  \bibinfo {series} {Encyclopedia of Mathematics and Its Applications, 124},
  Vol.~\bibinfo {volume} {1}\ (\bibinfo  {publisher} {Cambridge University
  Press},\ \bibinfo {year} {2009})\BibitemShut {NoStop}%
\bibitem [{\citenamefont {Chavez}\ \emph {et~al.}(2006)\citenamefont {Chavez},
  \citenamefont {Besserve}, \citenamefont {Adam},\ and\ \citenamefont
  {Martinerie}}]{chavez2006towards}%
  \BibitemOpen
  \bibfield  {author} {\bibinfo {author} {\bibfnamefont {M.}~\bibnamefont
  {Chavez}}, \bibinfo {author} {\bibfnamefont {M.}~\bibnamefont {Besserve}},
  \bibinfo {author} {\bibfnamefont {C.}~\bibnamefont {Adam}},\ and\ \bibinfo
  {author} {\bibfnamefont {J.}~\bibnamefont {Martinerie}},\ }\bibfield  {title}
  {\bibinfo {title} {Towards a proper estimation of phase synchronization from
  time series},\ }\href@noop {} {\bibfield  {journal} {\bibinfo  {journal}
  {Journal of Neuroscience methods}\ }\textbf {\bibinfo {volume} {154}},\
  \bibinfo {pages} {149} (\bibinfo {year} {2006})}\BibitemShut {NoStop}%
\bibitem [{\citenamefont {Fujisawa}\ and\ \citenamefont
  {Buzs{\'a}ki}(2011)}]{fujisawa20114}%
  \BibitemOpen
  \bibfield  {author} {\bibinfo {author} {\bibfnamefont {S.}~\bibnamefont
  {Fujisawa}}\ and\ \bibinfo {author} {\bibfnamefont {G.}~\bibnamefont
  {Buzs{\'a}ki}},\ }\bibfield  {title} {\bibinfo {title} {A 4 Hz oscillation
  adaptively synchronizes prefrontal, VTA, and hippocampal activities},\
  }\href@noop {} {\bibfield  {journal} {\bibinfo  {journal} {Neuron}\ }\textbf
  {\bibinfo {volume} {72}},\ \bibinfo {pages} {153} (\bibinfo {year}
  {2011})}\BibitemShut {NoStop}%
\bibitem [{\citenamefont {Schreglmann}\ \emph {et~al.}(2021)\citenamefont
  {Schreglmann}, \citenamefont {Wang}, \citenamefont {Peach}, \citenamefont
  {Li}, \citenamefont {Zhang}, \citenamefont {Latorre}, \citenamefont {Rhodes},
  \citenamefont {Panella}, \citenamefont {Cassara}, \citenamefont {Boyden}
  \emph {et~al.}}]{schreglmann2021non}%
  \BibitemOpen
  \bibfield  {author} {\bibinfo {author} {\bibfnamefont {S.~R.}\ \bibnamefont
  {Schreglmann}}, \bibinfo {author} {\bibfnamefont {D.}~\bibnamefont {Wang}},
  \bibinfo {author} {\bibfnamefont {R.~L.}\ \bibnamefont {Peach}}, \bibinfo
  {author} {\bibfnamefont {J.}~\bibnamefont {Li}}, \bibinfo {author}
  {\bibfnamefont {X.}~\bibnamefont {Zhang}}, \bibinfo {author} {\bibfnamefont
  {A.}~\bibnamefont {Latorre}}, \bibinfo {author} {\bibfnamefont
  {E.}~\bibnamefont {Rhodes}}, \bibinfo {author} {\bibfnamefont
  {E.}~\bibnamefont {Panella}}, \bibinfo {author} {\bibfnamefont {A.~M.}\
  \bibnamefont {Cassara}}, \bibinfo {author} {\bibfnamefont {E.~S.}\
  \bibnamefont {Boyden}}, \emph {et~al.},\ }\bibfield  {title} {\bibinfo
  {title} {Non-invasive suppression of essential tremor via phase-locked
  disruption of its temporal coherence},\ }\href@noop {} {\bibfield  {journal}
  {\bibinfo  {journal} {Nature communications}\ }\textbf {\bibinfo {volume}
  {12}},\ \bibinfo {pages} {1} (\bibinfo {year} {2021})}\BibitemShut {NoStop}%
\bibitem [{\citenamefont {Cohen}\ \emph {et~al.}(1999)\citenamefont {Cohen},
  \citenamefont {Loughlin},\ and\ \citenamefont {Vakman}}]{cohen1999ambiguity}%
  \BibitemOpen
  \bibfield  {author} {\bibinfo {author} {\bibfnamefont {L.}~\bibnamefont
  {Cohen}}, \bibinfo {author} {\bibfnamefont {P.}~\bibnamefont {Loughlin}},\
  and\ \bibinfo {author} {\bibfnamefont {D.}~\bibnamefont {Vakman}},\
  }\bibfield  {title} {\bibinfo {title} {On an ambiguity in the definition of
  the amplitude and phase of a signal},\ }\href@noop {} {\bibfield  {journal}
  {\bibinfo  {journal} {Signal Processing}\ }\textbf {\bibinfo {volume} {79}},\
  \bibinfo {pages} {301} (\bibinfo {year} {1999})}\BibitemShut {NoStop}%
\bibitem [{\citenamefont {Delprat}\ \emph {et~al.}(1992)\citenamefont
  {Delprat}, \citenamefont {Escudi{\'e}}, \citenamefont {Guillemain},
  \citenamefont {Kronland-Martinet}, \citenamefont {Tchamitchian},\ and\
  \citenamefont {Torresani}}]{delprat1992}%
  \BibitemOpen
  \bibfield  {author} {\bibinfo {author} {\bibfnamefont {N.}~\bibnamefont
  {Delprat}}, \bibinfo {author} {\bibfnamefont {B.}~\bibnamefont
  {Escudi{\'e}}}, \bibinfo {author} {\bibfnamefont {P.}~\bibnamefont
  {Guillemain}}, \bibinfo {author} {\bibfnamefont {R.}~\bibnamefont
  {Kronland-Martinet}}, \bibinfo {author} {\bibfnamefont {P.}~\bibnamefont
  {Tchamitchian}},\ and\ \bibinfo {author} {\bibfnamefont {B.}~\bibnamefont
  {Torresani}},\ }\bibfield  {title} {\bibinfo {title} {Asymptotic wavelet and
  gabor analysis: Extraction of instantaneous frequencies},\ }\href@noop {}
  {\bibfield  {journal} {\bibinfo  {journal} {IEEE transactions on Information
  Theory}\ }\textbf {\bibinfo {volume} {38}},\ \bibinfo {pages} {644} (\bibinfo
  {year} {1992})}\BibitemShut {NoStop}%
\bibitem [{\citenamefont {Bedrosian}(1963)}]{bedrosian1963product}%
  \BibitemOpen
  \bibfield  {author} {\bibinfo {author} {\bibfnamefont {E.}~\bibnamefont
  {Bedrosian}},\ }\bibfield  {title} {\bibinfo {title} {A product theorem for
  Hilbert transforms},\ }\href@noop {} {\bibfield  {journal} {\bibinfo
  {journal} {Proceedings of the IEEE}\ }\textbf {\bibinfo {volume} {51}},\
  \bibinfo {pages} {868} (\bibinfo {year} {1963})}\BibitemShut {NoStop}%
\bibitem [{\citenamefont {Leys}\ \emph {et~al.}(2013)\citenamefont {Leys},
  \citenamefont {Ley}, \citenamefont {Klein}, \citenamefont {Bernard},\ and\
  \citenamefont {Licata}}]{leys2013detecting}%
  \BibitemOpen
  \bibfield  {author} {\bibinfo {author} {\bibfnamefont {C.}~\bibnamefont
  {Leys}}, \bibinfo {author} {\bibfnamefont {C.}~\bibnamefont {Ley}}, \bibinfo
  {author} {\bibfnamefont {O.}~\bibnamefont {Klein}}, \bibinfo {author}
  {\bibfnamefont {P.}~\bibnamefont {Bernard}},\ and\ \bibinfo {author}
  {\bibfnamefont {L.}~\bibnamefont {Licata}},\ }\bibfield  {title} {\bibinfo
  {title} {Detecting outliers: Do not use standard deviation around the mean,
  use absolute deviation around the median},\ }\href@noop {} {\bibfield
  {journal} {\bibinfo  {journal} {Journal of experimental social psychology}\
  }\textbf {\bibinfo {volume} {49}},\ \bibinfo {pages} {764} (\bibinfo {year}
  {2013})}\BibitemShut {NoStop}%
\bibitem [{\citenamefont {Vitaterna}\ \emph {et~al.}(2001)\citenamefont
  {Vitaterna}, \citenamefont {Takahashi},\ and\ \citenamefont
  {Turek}}]{vitaterna2001overview}%
  \BibitemOpen
  \bibfield  {author} {\bibinfo {author} {\bibfnamefont {M.~H.}\ \bibnamefont
  {Vitaterna}}, \bibinfo {author} {\bibfnamefont {J.~S.}\ \bibnamefont
  {Takahashi}},\ and\ \bibinfo {author} {\bibfnamefont {F.~W.}\ \bibnamefont
  {Turek}},\ }\bibfield  {title} {\bibinfo {title} {Overview of circadian
  rhythms},\ }\href@noop {} {\bibfield  {journal} {\bibinfo  {journal} {Alcohol
  research \& health}\ }\textbf {\bibinfo {volume} {25}},\ \bibinfo {pages}
  {85} (\bibinfo {year} {2001})}\BibitemShut {NoStop}%
\bibitem [{\citenamefont {Gengel}\ and\ \citenamefont
  {Pikovsky}(2022)}]{gengel2022phase}%
  \BibitemOpen
  \bibfield  {author} {\bibinfo {author} {\bibfnamefont {E.}~\bibnamefont
  {Gengel}}\ and\ \bibinfo {author} {\bibfnamefont {A.}~\bibnamefont
  {Pikovsky}},\ }\bibfield  {title} {\bibinfo {title} {Phase reconstruction
  from oscillatory data with iterated Hilbert transform embeddings—benefits
  and limitations},\ }\href@noop {} {\bibfield  {journal} {\bibinfo  {journal}
  {Physica D: Nonlinear Phenomena}\ }\textbf {\bibinfo {volume} {429}},\
  \bibinfo {pages} {133070} (\bibinfo {year} {2022})}\BibitemShut {NoStop}%
\bibitem [{\citenamefont {Li}\ \emph {et~al.}(2010)\citenamefont {Li},
  \citenamefont {Li},\ and\ \citenamefont {Qian}}]{li2010discrete}%
  \BibitemOpen
  \bibfield  {author} {\bibinfo {author} {\bibfnamefont {H.}~\bibnamefont
  {Li}}, \bibinfo {author} {\bibfnamefont {L.}~\bibnamefont {Li}},\ and\
  \bibinfo {author} {\bibfnamefont {T.}~\bibnamefont {Qian}},\ }\bibfield
  {title} {\bibinfo {title} {Discrete-time analytic signals and Bedrosian
  product theorems},\ }\href@noop {} {\bibfield  {journal} {\bibinfo  {journal}
  {Digital Signal Processing}\ }\textbf {\bibinfo {volume} {20}},\ \bibinfo
  {pages} {982} (\bibinfo {year} {2010})}\BibitemShut {NoStop}%
\end{thebibliography}%

\section{Acknowledgements}
We thank Hiroshi Nishimaru, Shigeyoshi Fujisawa and Shiho Inagaki for helpful discussions. 
This study was supported by JSPS KAKENHI (No. 21J10799) to A.M.,  JSPS KAKENHI (No. 21K12056) to H.K., and JSPS KAKENHI (Nos. 18K11560, 19H01133, 21H03559, 21H04571, and 22H03695), JST PRESTO (No. JPMJPR1925), JST Moonshot R\&D (Grant Number JPMJMS2284), and AMED (No. JP21wm0525004) to R.K.

\section{Author contributions statement}
A.M, H.K., and R.K conceived the project. A.M. and R.K. developed the extended Hilbert transform method. A.M. performed numerical simulations and analyzed data. All the authors wrote the manuscript. H.K. and R.K. supervised the project. 

\end{document}